\def\BGP{boarding gate problem}
\def\DFBM{democratic fiber bundle model}
\documentstyle[aps,pre,epsf]{revtex}

\author{Rava da Silveira\\
Department of Physics, Massachusetts Institute of Technology,
Cambridge, MA  02139}
\title{An Introduction to Breakdown Phenomena in Disordered Systems}

\begin{document}

\maketitle
\begin{abstract}
The rupture of a medium under stress typifies breakdown phenomena. More
generally, the latter encompass the dynamics of systems of many interacting
elements governed by the interplay of a driving force with a pinning
disorder, resulting in a macroscopic transition.
A simple mean-field formalism incorporating these
features is presented and applied to systems representative of
fracture phenomena, social dilemmas, and magnets out of equilibrium. The
similarities and differences in the corresponding mathematical structures
are emphasized. The solutions are best obtained from a graphical method,
from which very general conclusions may be drawn. In particular, the various 
classes of
disorder distribution are treated without reference to a particular
analytical or numerical form, and are found to lead to qualitatively
different transitions. Finally, the notion of effective (or
phenomenological) theory is introduced and illustrated for
non-equilibrium disordered magnets.
\end{abstract}

\section{Introduction}

When boarding passengers, airline personnel usually call a limited
number of rows at a time --- ``Now seating rows 20 to 30!'' --- 
to ensure an efficient and orderly filling of the aircraft.
Nevertheless, there are always some hurried
individuals who unduly join the forming line before their row is called, only 
to be
rebuked by the flight attendant. In most instances the line is barely
disrupted,
while in some cases, such as delayed flights,
the ``avalanching'' of disobedient passengers creates a long, disorderly
line, absorbing nearly everyone. Why is this? Let us consider the mechanism by
which the line forms in greater detail. First, the forty or so passengers whose
rows have been called
move toward the gate, initiating a line. Next, a few ``impatients''
(or ``disobedients'') add to the ranks, and perhaps also some absent-minded
physicists who stand up as soon as they hear an announcement
and innocently approach the gate. Seeing this, some of the people
still sitting and waiting calmly, might lose their patience and
decide to join the growing line. This ``domino effect'' may die out
very soon, with just a few outsiders hiding in the queue, or it may
continue, resulting in a catastrophic boarding process.

The line formation at a boarding gate incorporates all the elements of
a {\it breakdown phenomenon} \cite{chakrabarti}:
it involves a large number of elements
(individuals), each in one of several possible states (sitting or in line),
which interact (the longer the line, the
stronger the temptation), and are driven as a whole (calls of the flight
attendant). Furthermore, different people have different ``obedience
thresholds'' that depend on their personality, mood, schedule,
etc. For a given length of the line, a fraction of these
thresholds are crossed and the resulting disobedients stand up and
move toward the gate. Equivalently, a sitting individual decides to
stand up and join the queue with some probability (dependent on the
line's length), thereby introducing an element of stochasticity, or
{\it disorder}. Similar features are present in a great variety of
processes, occurring, for example,
in materials under tension \cite{peirce,daniels,comment},
conductors with an imposed current \cite{chakrabarti,zapperi},
magnets driven by an applied field\cite{sethna}, and social
phenomena such as the ``boarding-gate problem.''
The rupture of the material, destruction of the conductor, macroscopic
flipping of the magnetization, or the failure of the social process,
respectively, can be viewed as a breakdown of the system.

The present paper aims to draw a simple quantitative picture of
breakdown phenomena in disordered systems, unifying
the various realizations into a single formalism while still
allowing for the details that distinguish each case. The most drastic
simplification underlying our approach is its ``mean-field'' nature,
which relies
on average quantities by ignoring fluctuations and, consequently,
any idea of locality or modulated spatial correlation. Although
such a crude description can depart significantly from three-dimensional
reality, it is often
accurate in higher dimensions and establishes a coarse framework which may
then be refined to include fluctuations (perturbatively). Nevertheless,
some systems, such as magnets with long-range interactions or the
boarding-gate line, are properly described by a mean-field theory;
in the \BGP, for example, any notion of locality is absent from the
interactions,  as any passenger can see the line as well as any
other.

In addition to differences intrinsic to various examples, our approach
should allow for the introduction of more than one type of disorder, as even
a given  system's behavior might very well depend on the character of the
stochasticity present. Better, we would like to divide all the
possible disorder distributions into classes, each associated with
a distinct qualitative behavior. To achieve this, we use
a graphical scheme\cite{daniels,comment} where we can {\it draw}
different  disorder
distributions, enabling us to handle the latter in a very general fashion,
without reference to a particular analytical or numerical form, and to
distinguish the features relevant to the breakdown
phenomenon\cite{comment}.  These
features define the various ``universality classes,'' identified by one of 
their
(drawn) members or by a generic statement (for example, ``the class
of all  {\it continuous} distributions''). A graphical method also
avoids the possible misconceptions of a more specific
treatment\cite{andersen}. Indeed, if the qualitative behavior of a
physical system changes upon making a particular, for example,
Gaussian, distribution wider, what is the associated key feature?
Is it the variance of the Gaussian or its maximum value? Finally, a
graphical scheme allows us to ``follow'' the state of the system as
it is driven, thus capturing the whole breakdown phenomenon in a
single picture.

We develop the method in the context of
fractures (Sec.~II), for which it was originally
devised\cite{peirce,comment}, before applying it to two other
realizations interesting in their own right. In Sec.~III, some
quantitative results are obtained for the \BGP. Section~IV focuses
on driven magnets, a workhorse of non-equilibrium statistical
mechanics as well as a central model for hysteresis. In Sec.~V, we
introduce the notion of effective (or phenomenological) theory, a
broader framework than the models examined. We return to the
discussion of driven magnets in this more general context by
constructing a simple effective theory suited to the problem, and
mention some natural extensions to richer, non-mean-field
approaches. Finally, we conclude (Sec.~VI) with a summarizing
picture and a brief allusion to dynamics.

\section{Fractures and Earthquakes: the Democratic Fiber Bundle Model}

The democratic fiber bundle model was introduced by
Peirce\cite{peirce} in 1926, and formalized and studied in greater
detail by Daniels\cite{daniels} in 1945. It consists of $N_0$
($\rightarrow \infty$) fibers pulled by a force $F$ which they
share uniformly (democratically). As the force is increased, a
fiber breaks if the stress it undergoes exceeds a threshold (or
strength), drawn from a distribution $p(x)$. After each set of
failures, the total force is redistributed over the remaining
fibers. As members of the {\it British Cotton Industry Research
Association} and the {\it Wool Industries Research Association},
respectively, Peirce and Daniels sought a more fundamental
understanding of the `hank' and `lea' tests for wool and cotton
yarns, in which a hank (or bundle) is stretched between two hooks
until it ruptures. A more life-threatening illustration of the
model is offered by the failure of an elevator rope (Fig.~1).

\begin{figure} [h]
\epsfxsize=6truecm 
\vspace*{.7truecm} 
\centerline{
\epsfbox{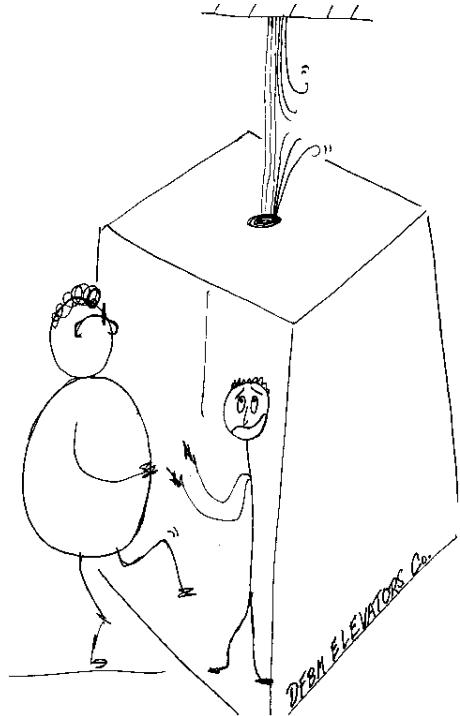} 
} 
\vspace*{.3truecm}
\caption{The democratic fiber bundle model realized by an elevator rope.}
\vspace*{.5truecm}
\end{figure}

The \DFBM\ has further been applied to the
study of cracks and fractures\cite{andersen} and
earthquakes\cite{sornette}, as a simplified model that retains some
of the essential features of the original problem. In a failing
medium, stress is continually redistributed due to the nucleation,
growth, and coalescence of cracks. Realistic descriptions  of
fractures and earthquakes may lie anywhere between this democratic
or infinite-range limit embodied by the \DFBM, also sometimes
referred to as the ``global-load-sharing'' or
``equal-load-sharing'' model, and the local or short-range limit,
the ``local-load-sharing'' model\cite{wu}, in which the stress
released by a broken bond is passed on to its immediate (intact) neighbors.
Far from being
one-dimensional or democratic, real-world systems may be (effectively) two- 
or three-dimensional,
with connectivity properties that lead to a more complicated and
richer picture. Alluding to the applicability of the model to the
testing of a cloth sample rather than a hank, Daniels himself
recognized\cite{daniels} that ``in closely woven fabrics the cross-threads afford
a measure of support which introduces complications.''

Beyond the obvious question of the bundle's strength (maximum sustainable
stress without full rupture) addressed by Peirce\cite{peirce} 
and Daniels\cite{daniels}, we
would like to uncover, in the large $N_0$ limit, its behavior
characterized by the number $N(F)$ of intact fibers as
the force $F$ increases from $0$ to $\infty$. In particular, we
investigate the critical properties of the fiber bundle upon
approaching a macroscopic failure, and the generic classes in which
they fall.

Upon incrementing the force from $F-dF$ to $F$, a few fibers whose
thresholds have been overshot, break. As a result, the stress increases on the
remaining intact fibers, leading to a secondary set of failures. This
triggers a tertiary set of failures, and so on, until the bundle stabilizes
at a non-vanishing $N(F)$ or snaps off as a whole
($N(F) =0$). Consider the $i$th rupture which leaves
$N_i$ unbroken fibers, each now under a stress $F/N_i$. The
thresholds of a fraction of intact fibers have been exceeded,
namely $\int_{F/N_{i-1}}^{F/N_i}p(x) dx$, 
and the next set of failures 
brings their number down to 
\begin{equation}
\label{iterationN}N_{i+1}=N_i-N_0\int_{F/N_{i-1}}^{F/N_i}p(
x) dx=
N_0\left\{ 1-P\left( \frac F{N_i}\right) \right\} ,
\end{equation}
where $P$ is the cumulative distribution defined by 
\begin{equation}
\label{cumulative}P(x)
=\int_0^xp(x^{\prime}) dx^{\prime}.
\end{equation} 
Clearly, the function
$N(F)$, defined as the number of fibers remaining
unbroken under a total force $F$,
is none other than $N_\infty$ calculated at $F$ (the limiting value
of $N_i$ as $i\rightarrow\infty$).

Problems stated in terms of a recursion relation often have an elegant
graphical solution which displays in a single picture the presence or
absence of convergence. Here, a graphical
scheme for the iteration is best constructed with the quantities
$x_i=N_i/F$,
$f=F/N_0$, and 
\begin{equation}
\label{pi}\pi(x) =1-P\left( \frac 1 x\right),
\end{equation} 
in terms of
which Eq.~(\ref{iterationN}) becomes 
\begin{equation}
\label{iterationX}fx_{i+1}=\pi(x_i) .
\end{equation}
The function $\pi(x)$ increases monotonically from $0$
at $x=0$ to 1 at $x=\infty$ and a graphical iteration of
Eq.~(\ref{iterationX})
shows that $N(F) =N_\infty$ is given by the right-most
intersection of the curve $y=\pi(x)$ with the straight
line $ y=fx$. As the force is increased, the straight line becomes
steeper and the abscissa $N(F)/F$ of the
intersection point moves to the left. Figure~2 illustrates the
graphical scheme for three successive increments of the force.
First, $f$ is switched from 0 to $f_1$; for $f\rightarrow0$, the
straight line is flat and intersects $\pi(x)$ at
$x=\infty$ (as expected from
$x_0=N_0/F$) and $y=1$. After the sudden switch, fibers with a threshold
smaller than $f_1$ immediately break, causing more ruptures according to
Eq.~(\ref{iterationX}). The corresponding iterations are
represented by dotted
arrows, and terminate at the intersection point with
abscissa $x_\infty\left(f_1\right)$. Next, $f$ is switched
from $f_1$ to $f_2$, and similarly, $x_i$ converges to
$x_\infty\left(f_2\right)$. Finally, $f$ is switched to $f_3$, and the
iterations lead all the way to $x=0$, which corresponds to a fully
ruptured bundle. If the force, rather than suddenly switching from $f_1$
to $f_2$, goes through several intermediary steps, $x$ successively
converges to the abscissa of the corresponding intersection points, but
again ends up at $x_\infty(f_2)$ when $f=f_2$. Thus
$N(F)$ is a well defined function of $F$, independent of the
past history of the force as long as it is increased monotonically. 

\begin{figure}
\epsfxsize=8truecm 
\vspace*{.7truecm} 
\centerline{
\epsfbox{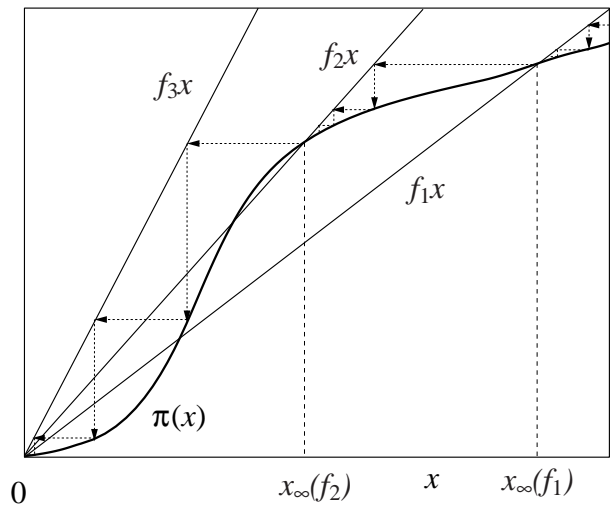}
} 
\vspace*{.3truecm}

\caption{Illustration of the graphical scheme for three successive 
increments in the force. The dotted arrows represent iterations of 
Eq.~(\ref{iterationX}). The latter terminate, for each value of the
force, at the right-most intersection, yielding the number of intact
fibers as $N(F)=Fx_\infty$.}
\vspace*{.5truecm}
\end{figure}

In what follows, we
focus on a continuous steady increase of $F$, which can be thought of as
the limit of many small successive increments. We assume that the
iterations occur fast enough, or equivalently, that $F$ varies slowly
enough for the convergence to be considered immediate. By analogy to
thermodynamical processes, we may say that the force is increased {\it
quasi-statically}.

We are now in a position to investigate the critical behavior of the
\DFBM\ with various threshold distributions $p(x)$. The
graphical method provides generic conclusions, based on a (drawn)
arbitrary curve $y=\pi (x)$. Thus the results are
independent of a specific analytical or numerical form for
$p(x)$, but follow from the qualitative features of the
curve. The latter identify a class of distributions, associated
with a given behavior of the bundle. For continuous distributions,
there are {\it three generic classes}\cite{comment}, as
illustrated in Fig.~3. These classes  are distinguished by the
character of $p\left(x\right)$ at large argument, which dictates the bundle's 
behavior
under large forces, through the slope
\begin{equation}
\label{derivative}\frac{d\pi(x)}{dx} =\frac 1{x^2}\,p(
\frac 1x) .
\end{equation}

\begin{figure} 
\epsfxsize=8.1truecm 
\vspace*{.5truecm} 
\centerline{
\epsfbox{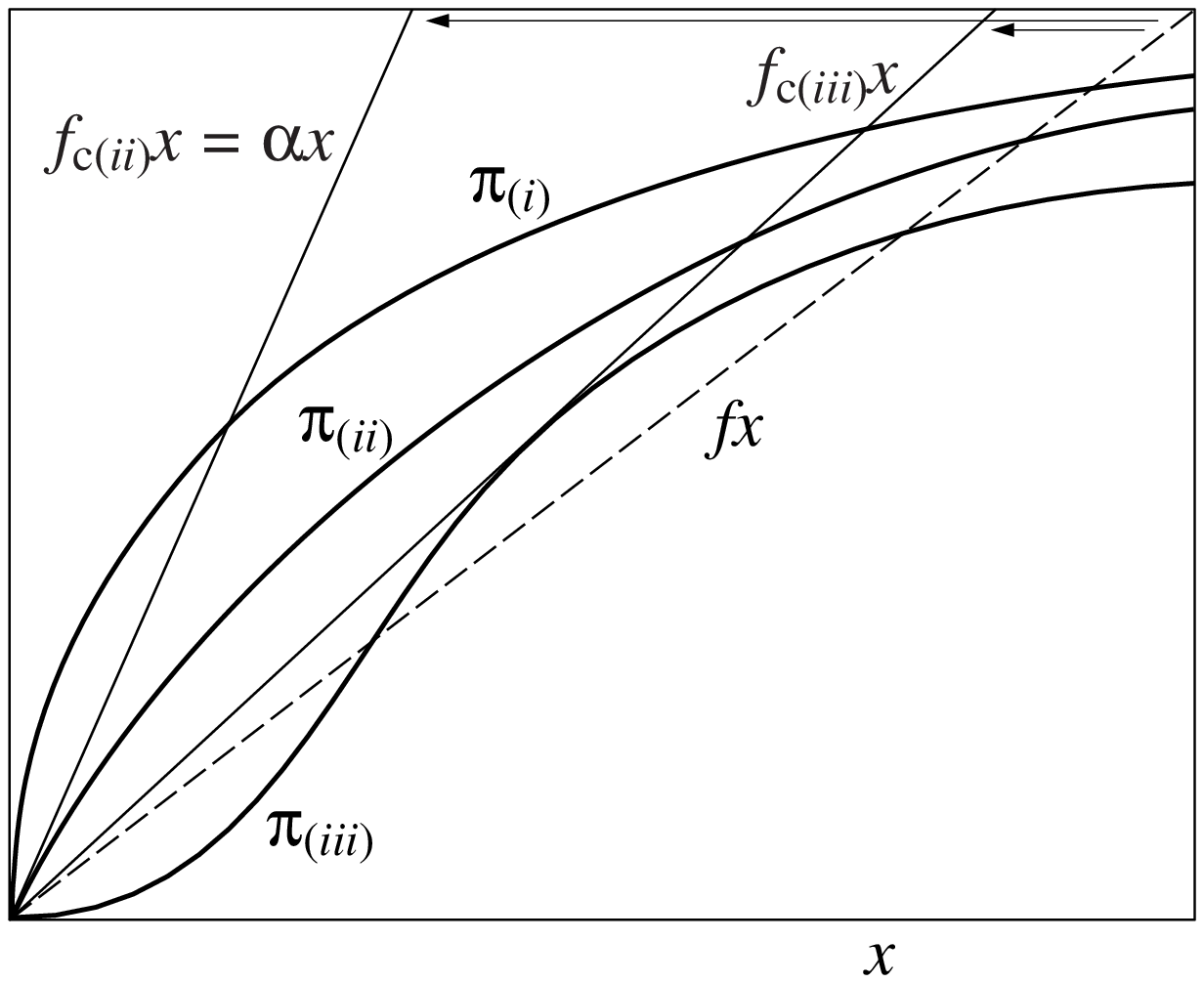} 
} 
\end{figure}

\begin{figure}
\epsfxsize=8.7truecm 
\vspace*{.5truecm} 
\centerline{
\epsfbox{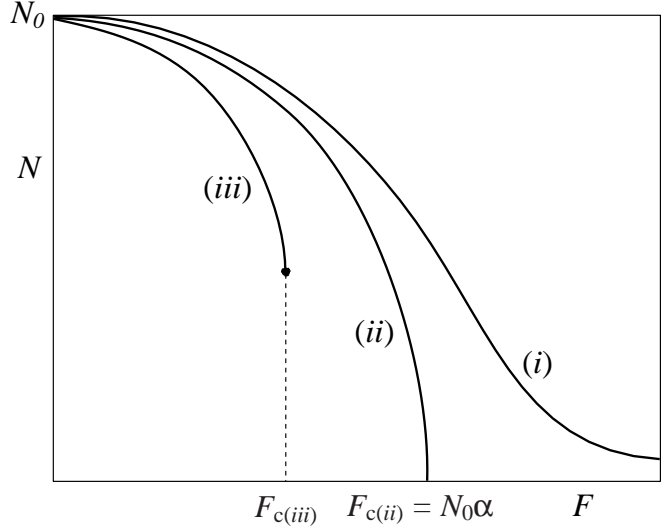} 
} 
\vspace*{.3truecm}

\caption{The democratic fiber bundle model with continuous threshold 
distributions. (a) Illustration of the graphical scheme. As the
force is increased, the straight line becomes steeper and the
abscissa of the intersection points, proportional to the number of
intact fibers, move to the left. Typical solutions are shown in (b).
Intact fibers are always left in bundles of type ({\it i}), which
never rupture completely. Bundles of type ({\it ii}) fail
continuously at a force 
$F_{\mbox{c}(ii)}=N_{0}\alpha$, whereas bundles of type ({\it iii}) 
fail discontinuously at $F_{\mbox{c}(iii)}$.}
\vspace*{.5truecm}
\end{figure}

\noindent
{\it Class} ({\it i}). For $p(x) \sim x^{-r}$
($x\rightarrow 
\infty$) with
$r<2$ (normalization requires $r>1$), that is, distributions with an
appreciable fraction of robust fibers, the bundle never fails completely.
Under any force $F$, some intact fibers are left.

\noindent
{\it Class} ({\it ii}). In the limiting case where $p(x)
=\alpha  x^{-2}$
as $x\rightarrow \infty$,
the number of intact fibers goes continuously to zero at
$F_c=N_0\alpha$. Furthermore, the response function of the bundle
or  {\it breaking rate} $dN/dF$ diverges as
$\left(F_c-F\right)^{-1/s}$, where $s$ is the subleading
power entering the distribution function, $p\left(x\right) =\alpha
x^{-2}+\beta x^{-s}+\ldots,\ s>2$. Following the nomenclature
adopted in the field of critical phenomena, we may refer to the
macroscopic rupture of the bundle associated with a diverging
response function as a second-order transition with exponent
$\gamma =1/s$.

An example of a ``fat-tail distribution'' is $p(x)
=(\alpha/x^r)\, e^{-\alpha /x^{r-1}}$ for any $\alpha$, with
$1<r<2$ in class ({\it i}) and $r=2$ (and exponent $\gamma =\left(
2r-1\right)^{-1}=1/3$) in class ({\it ii}).

\noindent
{\it Class} ({\it iii}). For any narrower distribution, such that
$x^2p(x)
\rightarrow 0$ for $x\rightarrow \infty$, there is a macroscopic
failure of the bundle at a finite critical force $F_c$ at
which the number of intact fibers falls discontinuously from
$N(F_c)$ to 0. The breaking rate diverges
in general as
\begin{equation}
\frac{dN}{dF}\sim \left( F_c-F\right)^{\gamma}, \quad
\gamma=-\frac1 2.
\end{equation}
A ``trivial'' exponent of $\gamma=-1/2$ is to be expected within a mean-field 
approach which
ignores the effect of spatial fluctuations (see Sec.~V).
Of course, it is
possible to choose a particular distribution with a ``bump'' such that $
\gamma =1/2n$, with $n$ an integer, instead of $\gamma =1/2$, but it
reflects a fine-tuned choice rather than the generic one. All continuous
distributions with finite mean, subsumed in class ({\it iii}), lead to a
macroscopic failure of the bundle at a finite critical force, signaled by a
diverging breaking rate. Although this divergence in the response is still
reminiscent of a second-order transition, the situation differs from the
usual case where the order parameter (here $N(F)$) 
is continuous at the singularity. To
avoid any confusion, we refer to ruptures of type ({\it iii}) as 
{\it soft} failures. 

Two examples of distributions which belong to class
({\it iii}) are $p(x)
=\left(x/\lambda^2\right)e^{-x/\lambda}$  and 
$p\left( x\right) =\left(x/\lambda\right) e^{-x^2/2\lambda }$.
Other examples  include
finite-support distributions, which are non-vanishing only in an interval $
{\cal I}=\left[x_0,x_0+\lambda \right]$, provided $p(x)$
is
continuous on ${\cal I}$ and $p(x_0) =p(x_0+\lambda) =0$.

Fig.~3(b) illustrates the generic histories $N(F)$ for
each of the three classes. Clearly,
by having more ``bumps'' ($\pi^{(iii)}(x)$ on Fig.~3(a)
has one), that is, variations in the convexity of $\pi(x)$, the
bundle may undergo additional macroscopic (finite fraction of
$N_0$) failures at smaller forces. None of these, however, may lead
to a complete rupture, which always corresponds to one of the three
classes described above.

\begin{figure}
\epsfxsize=8truecm 
\centerline{
\epsfbox{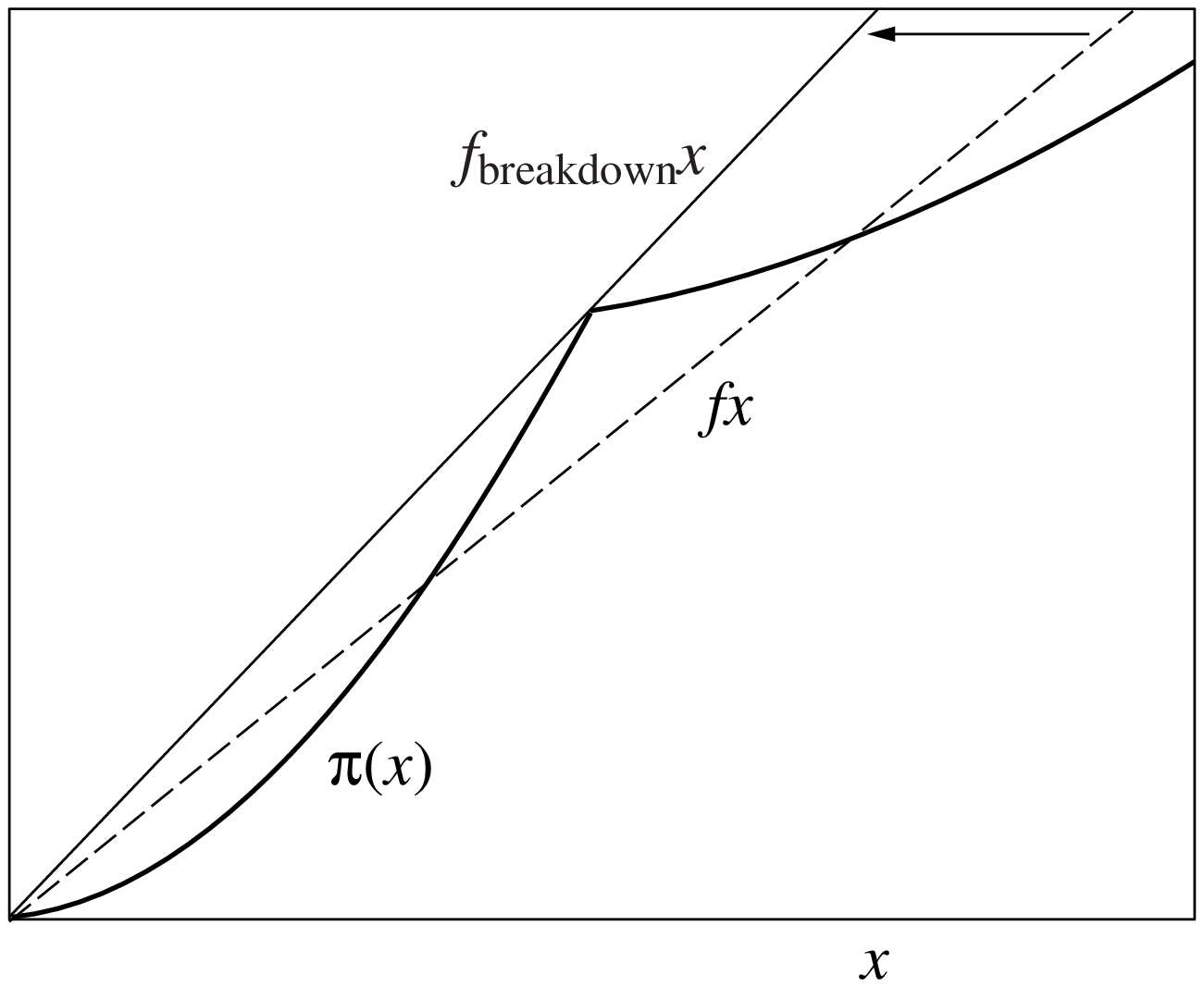} 
} 
\epsfxsize=8.6truecm 
\vspace*{.6truecm} 
\centerline{
\epsfbox{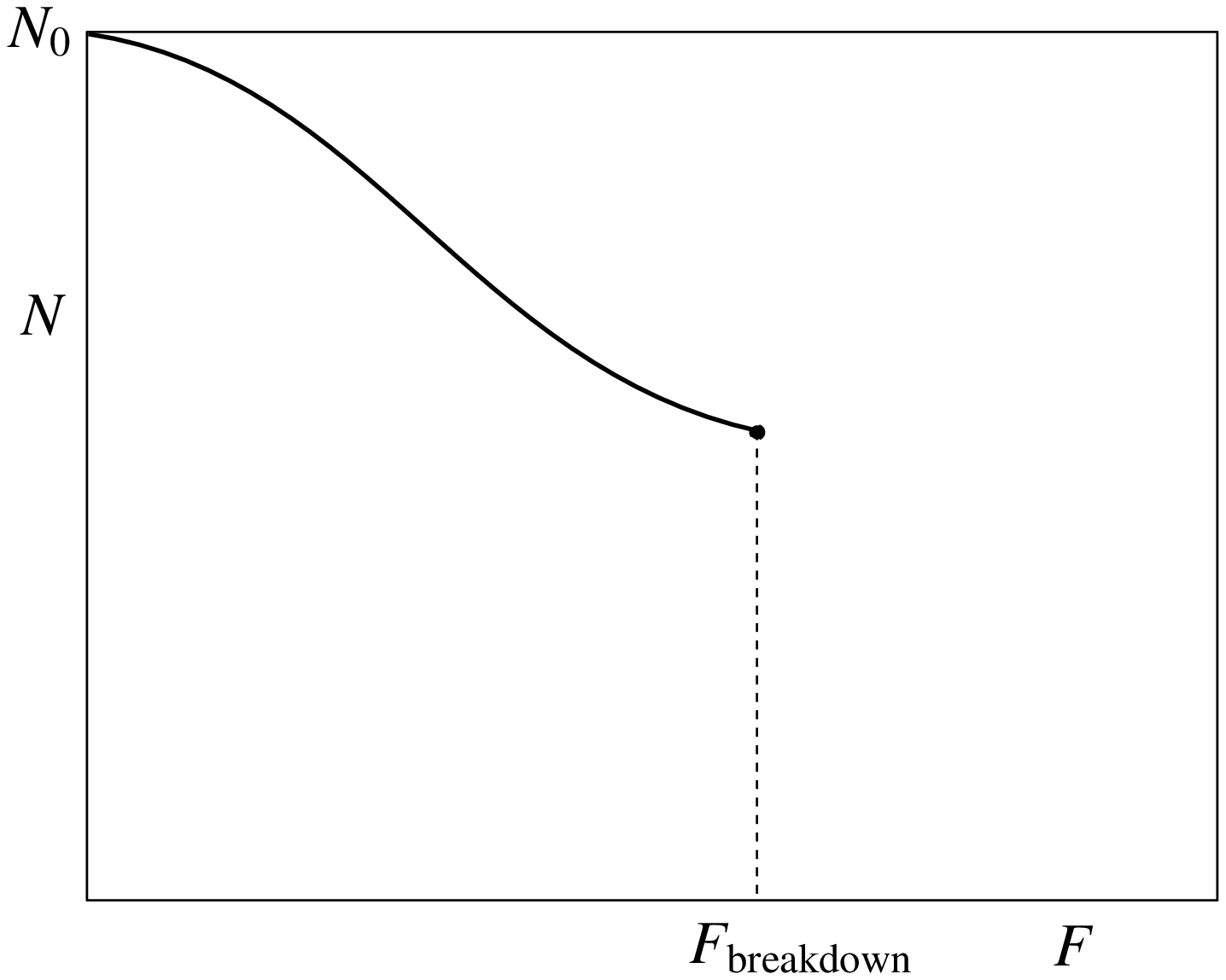} 
} 
\vspace*{.4truecm}

\caption{The democratic fiber bundle model for a markedly
discontinuous distribution of the thresholds. In (a), the graphical
scheme leads to (b), an abrupt failure of the bundle.}
\vspace*{.6truecm}
\end{figure}

The examples given for class ({\it iii}) are valid for any value of the
parameter $\lambda $. Thus, the threshold distribution may be as narrow
and peaked ($\lambda \rightarrow 0$) as desired and still lead to a soft
failure, provided $p(x)$ is continuous. An {\it abrupt}
rupture of  the bundle
{\it with no divergence in the breaking rate} preceding it is possible only
if $\pi(x)$ has a non-differentiable point 
(Fig.~4)\cite{comment}, which in
turn requires a discontinuity in $p$. Thus, abrupt failures in this
mean-field model, far from being a generic feature associated with narrow
disorder distributions, are an artifact of a singular point in $p$.
The phenomenon is in fact even more particular, because the {\it
magnitude} of the discontinuity matters. In the case of a
finite-support distribution on ${\cal I}$ with $p(x_0)
\neq 0$, for example, no fiber breaks up to a force $F_0=N_0x_0$.
At $F_0$, an abrupt failure occurs only if $d\pi
\left(1/x_0\right)/dx
\geq x_0$, as can be shown by the graphical method, which requires
a minimal jump $p(x_0) \geq 1/x_0$. Whether this
macroscopic failure yields full rupture of the bundle  or not
depends on the detailed form of $p$. If it does not, and if
$p$ is continuous on ${\cal I}$, the global bundle failure is soft
and belongs to one of the three cases examined above (case ({\it
iii}) if 
$\left\langle x\right\rangle =\int_{0}^{\infty} dx\;x\;p(x) 
<\infty $). The description of
a large $N_0$ bundle by a continuous distribution $p$ would seem more
physical, unless it is made of fibers of a few different types, each with a
well defined threshold. Consider for example a ``bimodal'' bundle with $\rho 
N_0$ ($0\leq \rho \leq 1$) fibers of strength $f_1$ and $(1-\rho)\,
N_0$ fibers of strength $f_2<f_1$\cite{daniels}. It is easy to see
from the graphical method (Fig.~5) that if
$f_2/f_1>\rho $, the entire bundle snaps off abruptly at
$F_2=N_0f_2$. On the other hand, if $f_2/f_1<\rho $ it undergoes an
abrupt failure at $F_2$ leaving the $\rho N_0$ stronger fibers
intact, and then ruptures completely (again abruptly) when the
force reaches 
$F_1=N_0f_1$.

\begin{figure} 
\epsfxsize=8truecm 
\vspace*{.3truecm} 
\centerline{
\epsfbox{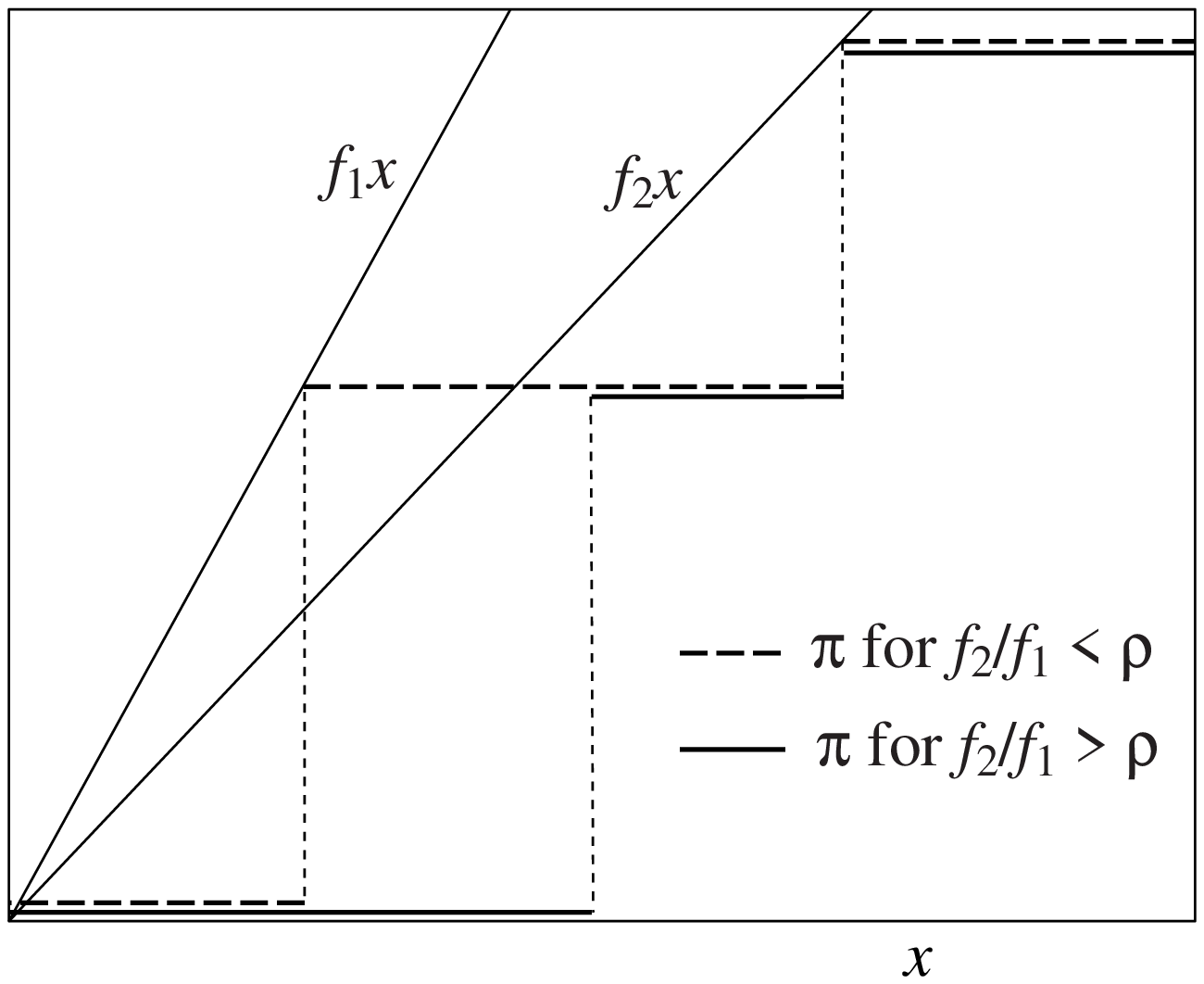} 
} 
\epsfxsize=8.9truecm 
\vspace*{.8truecm} 
\centerline{
\epsfbox{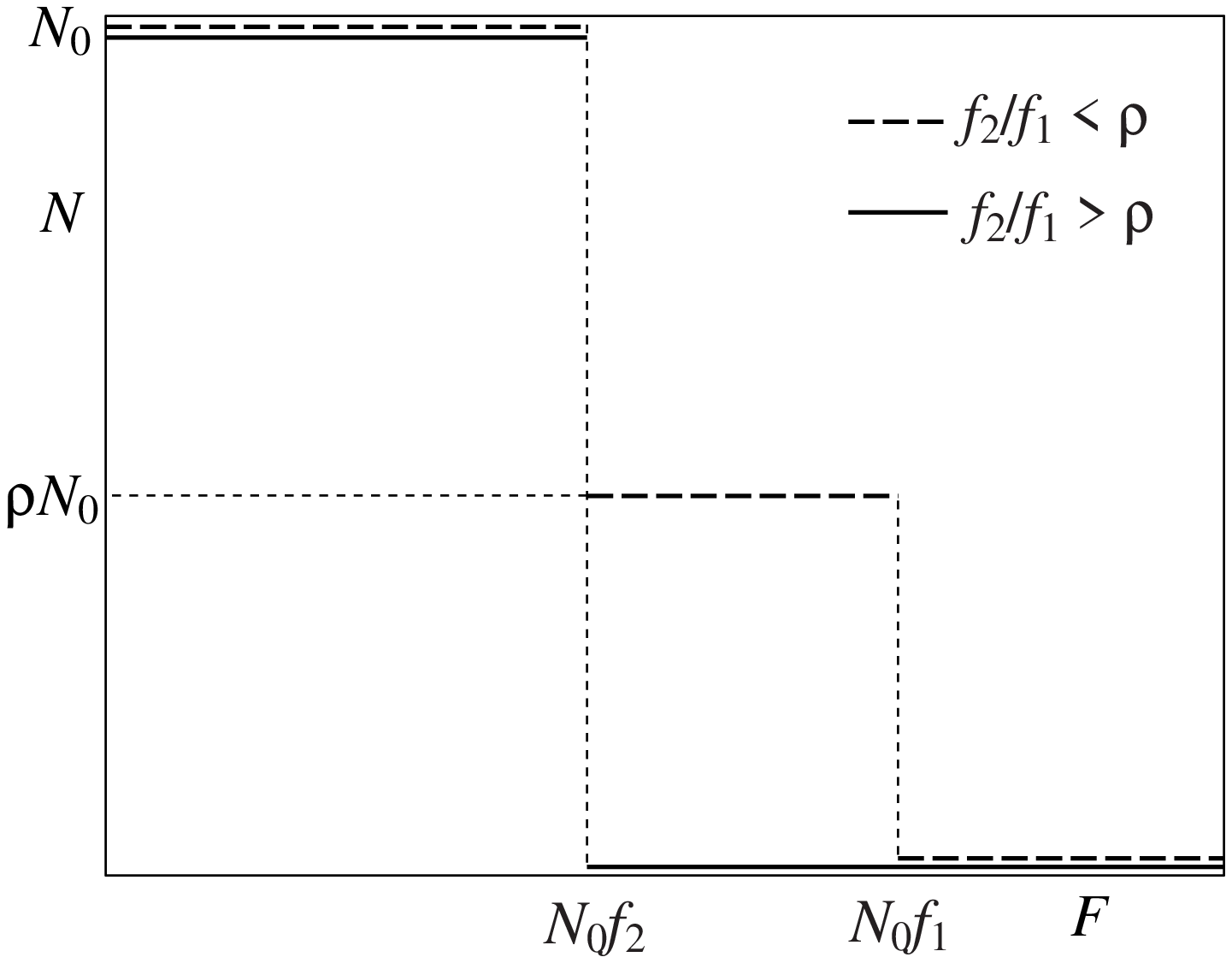} 
} 
\vspace*{.5truecm}

\caption{The democratic fiber bundle model for a ``bimodal''
distribution consisting in $\rho N_{0}$ fibers of strength $f_{1}$ and 
$\left(1-\rho\right)N_{0}$ fibers of strength $f_{2}$. 
(a) Illustration of the graphical scheme and (b) typical solutions. The bundle
fails in either one or two abrupt steps.}
\vspace*{.7truecm}
\end{figure}

A physical medium which fails as a result of the growth of a single large
crack is often referred to as ``ductile,'' in analogy with flexible
materials, such as most metals, which rearrange under stress and can thus be 
pulled into long thin
threads. At the other end of the spectrum, fragile or ``brittle'' media are
characterized by a sudden failure due to the coalescence of many
microcracks. In the context of our mean-field model, we may identify abrupt
failures from markedly discontinuous distributions (such as narrow
finite-support ones or bimodal ones) with brittleness, and soft failures
from the continuous distributions of classes ({\it ii}) and ({\it iii}) with
ductility. Thus the character of a
medium's strength depends on the {\it shape} of the distribution of
randomness. 

The failure of an element, whether a fiber in an elevator rope
or a patch of a tectonic plate, is often accompanied by the emission of a
sound\cite{garciamartin}. Before an abrupt failure, nothing unusual
is heard,  whereas an
increasingly loud noise is produced by a soft failure. Thus impurities,
dirt, and disorder might be salutary in times of calamity!

\section{Social Logistics: The Boarding Gate Problem}

Armed with the method developed in Sec~II, we return to
the boarding-gate problem of the introduction, and treat it in a
quantitative fashion. The similarity between the breaking of a fiber bundle
and the growth of a boarding-gate line becomes more apparent through the
following mapping of the degrees of freedom, disorder, driving agent, and
coupling:
$$
\begin{array}{ccc}
\text{sitting person, waiting} & \longleftrightarrow & \text{intact fiber,}
\\ \text{standing person, in line} & \longleftrightarrow & \text{broken
fiber,} \\ \text{``obedience threshold''} & \longleftrightarrow &
\text{ strength threshold,} \\ \text{flight attendant's calls} & 
\longleftrightarrow & \text{driving force,} \\ \text{temptation from seeing
the line} & \longleftrightarrow & \text{redistribution of the force.} 
\end{array}
$$
Nevertheless, the two problems are different in that the stress applied to
any one intact
fiber is inversely proportional to their total number, whereas the
temptation here grows with the length of the line. Furthermore, for
simplicity we focus on a single call of the flight attendant, say the first
one. In the language of the \DFBM, this corresponds to switching
the force from zero to some value, and leaving it there. Finally,
the ``thermodynamic limit'' we treat (in implicitly assuming
infinitely many passengers) may be a  poor
approximation for a situation involving a few dozen or a few hundred people.
The fact that the first group of passengers called, and consequently the
ones still waiting, may not be representative of the overall threshold
distribution, for example, is one of the possibly important finite-size
effects neglected.

Let us say that the flight attendant first calls $\ell$ passengers, who
readily form a line. Seeing the latter, a few disobedients join in, equal in
number to $(N-\ell) P(\ell)$, where $N$ is the
total number of passengers waiting to board the flight and $P$ is
the cumulative  threshold
distribution. More precisely, $p(x) dx=(dP/dx)dx$ is
the probability that an individual unduly joins the line, when it
reaches a length between $x$ and $x+dx$. The process might stop at
this point, leaving the few intruders in line, or
might keep on ``avalanching.'' In general, we can write a recursive
relation describing the growth of the line, analogous to Eq.~(\ref
{iterationN}), as 
\begin{equation}
\label{iterationBGP}n_{i+1}=\ell+(N-\ell) P(n_i) 
\text{,}
\end{equation}
where $n_i$ is the number of people in line after $i$ iterations
($n_0\equiv\ell$). The avalanching
terminates at the left-most intersection of the line
$(n-\ell)/(N-\ell)$  with
the curve $P(n)$ and the eventual
length of the line, $n_\infty$, is read off as the abscissa of the
intersection point. In other words, the length of the line
is obtained as the smallest solution to 
\begin{equation}
\frac{n-\ell}{N-\ell}=P(n) , 
\end{equation}
leading to the following possible scenarios.

\noindent $\bullet$ Most optimistically, no one's threshold is
lower than the length of the initial line, that is, $p(x\leq \ell)
=0$ (Fig.~6).
No unexpected passenger shows up in the line, and the boarding
runs smoothly.

\noindent $\bullet$ More generally, for a large airplane with
hundreds of passengers, we expect thresholds to be widely
distributed. We consider distributions with $p(x) \neq
0$ for all $x\in \left[ 0,N\right]$ and $P(N) =1$,
according to which there are always some disobedients who join the
ranks no matter how small the temptation and no one is restrained
enough to stay seated when the line includes nearly all  the
passengers. In this case, the behavior of the line's growth is governed by the
character of $p$ at large argument (Fig.~7), reminiscent of the solution of 
the \DFBM\
with continuous distributions. If $p(N) <1/N$, the
line may anarchically absorb all passengers. If $p(N) >1/N$,
on the other hand, it must stabilize at $\ell<n_\infty <N$, the
precise value of which depends on the detailed form of the
function $p$.

\begin{figure}
\epsfxsize=9.2truecm 
\centerline{
\epsfbox{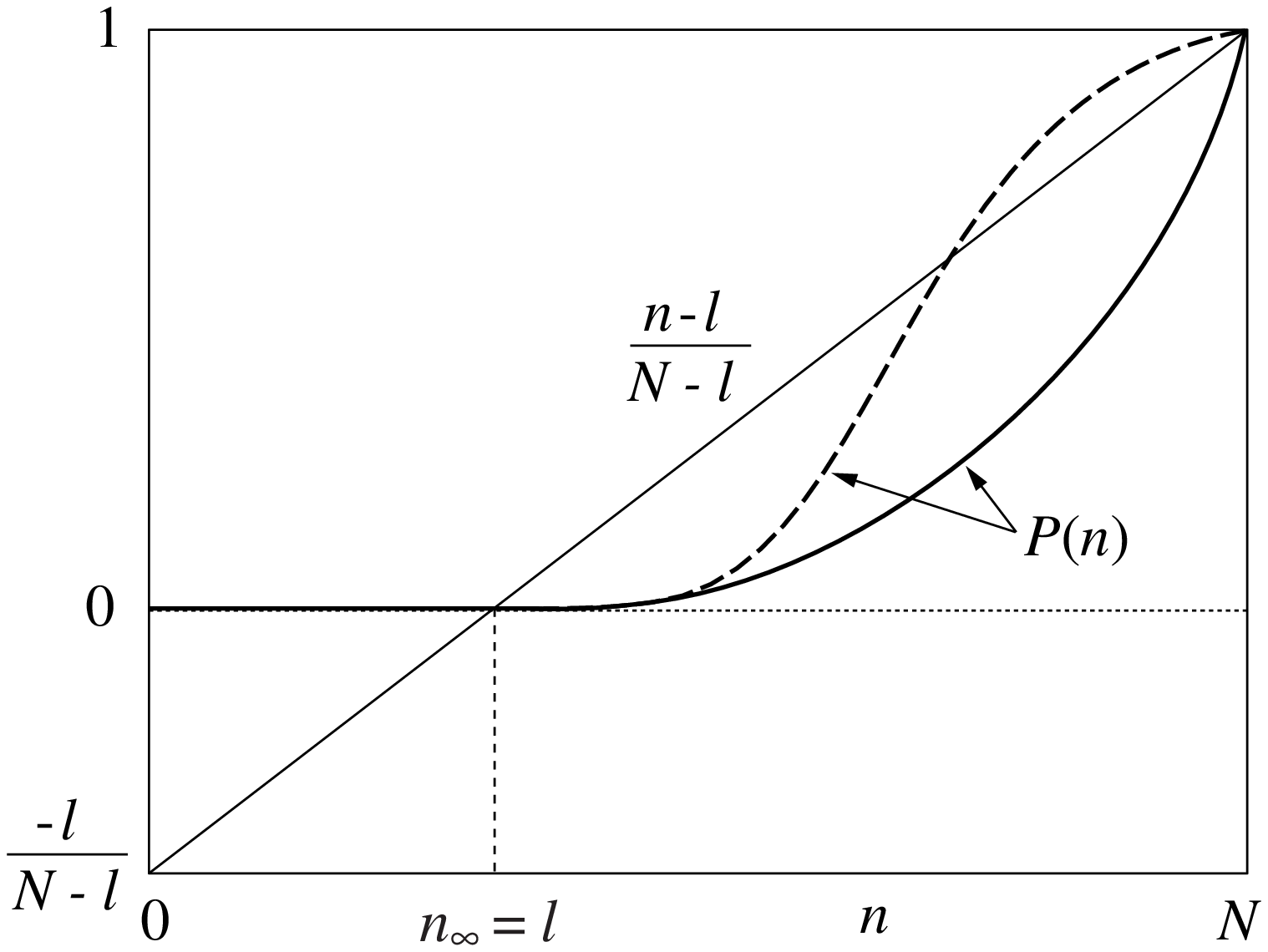} 
} 
\vspace*{.2truecm}

\caption{Graphical scheme for the boarding-gate problem with a threshold 
distribution
vanishing below $\ell$. The solid and dashed curves represent two possible cases,
both leading to lines with no intruders.}
\vspace*{1.2truecm}

\epsfxsize=9.2truecm 
\centerline{
\epsfbox{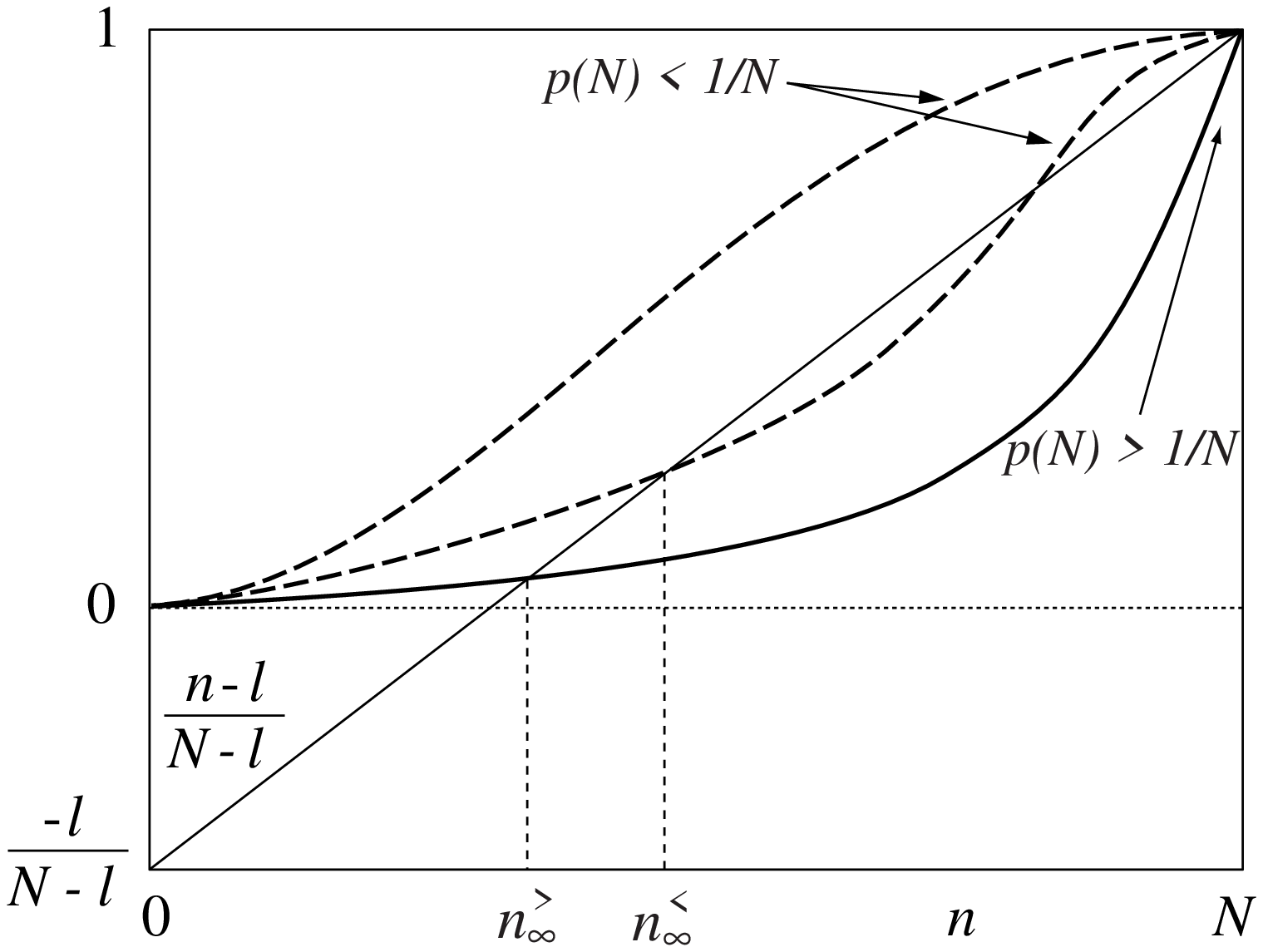} 
} 
\vspace*{.2truecm}

\caption{Graphical scheme for the boarding-gate problem with 
non-vanishing distributions. If $p(N)>1/N$ (solid line), 
$P(n)$
must cross the straight line $\left(n-\ell\right)/\left(N-\ell\right)$ at some 
$n$
smaller than $N$. If $p(N)<1/N$ (dashed lines), it may or may
not.}
\vspace*{.6truecm}
\end{figure}

\noindent $\bullet $ Often, there is a small fraction $n_p/N$
of {\it complete} disobedients,
a small fraction $n_s/N$ of {\it 
complete} obedients,
and the rest of the passengers'
thresholds scattered in between. For such a contingency, the
distribution reads
\begin{equation}
p(x) =\frac{n_p}N\delta(x) +g(x)
+\frac{n_s}N\delta(x-N) , 
\end{equation}
where $g(x)$ is a smooth function which, in a realistic
model, is appreciable between two bounds $n_1$ and $n_2$. Then, the
growth of the line saturates at a length
$\ell+\left(\frac{N-\ell}N\right) n_p<n_\infty <N-\left(
\frac{N-\ell}N\right) n_s$, which depends on the relative values of
$\ell$, $n_p$, and $n_1$ (Fig.~8). If
$n_1>\ell+n_p$, the eventual line includes few unexpected
passengers except the physicists --- a situation easy to cope with,
due to the small number of physicists. However, if
$n_1<\ell+n_p$ (and in particular if
$n_2-n_1
$ is small ($g$ very peaked)), the few physicists drag flocks of
others along, resulting in a line with all the passengers but the
complete obedients and perhaps a few highly obedient individuals.

\begin{figure} [h]
\epsfxsize=11.3truecm 
\vspace*{.2truecm} 
\centerline{
\epsfbox{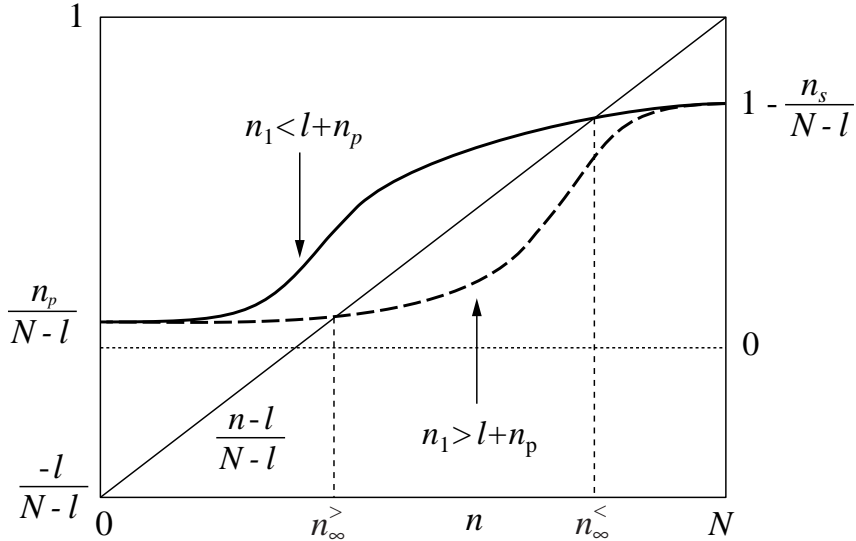} 
} 
\vspace*{.3truecm}

\caption{Graphical scheme for the boarding-gate problem with a distribution
reflecting a fraction of complete disobedients, a fraction of 
complete obedients, and the rest of the passengers peaked in between.
The character of the solution depends on whether the peak is closer to the
disobedient (solid line) or to the obedient (dashed line) side.}
\label{}  
\vspace*{.6truecm}
\end{figure}

There is no doubt that social phenomena involve a great many
parameters making them highly complex in comparison to ``simple''
physical systems such as a fiber bundle. However, in much the same
way as the breaking of a wool hank is faithfully described without
recourse to the molecular structure of the individual threads, a
{\it coarse} picture of some human phenomena may be drawn while
ignoring intricate psychological and other processes that each
individual goes through.  And it is amusing that such a {\it
phenomenological} approach yields a number of quantitative,
testable results for the behavior of a human group.

\section{Magnets Out of Equilibrium: Critical Hysteresis in a Driven Ising 
Model}

In the most modest speck of dust, an incredible number of electrons are
dancing around, creating small currents and with them magnetic moments.
Each electron also carries its own spontaneous moment, and it is the
complicated addition of all these effects which determines the total 
magnetization of the
dust. In the spirit of foregoing intricate details in favor of a coarser
but simpler picture, Ising and Lenz proposed a model\cite{ising} in
which a material is divided into many elements (labeled by their
position ${\bf  r}$), each
carrying a (unit) magnetic moment, or {\it spin}, which tends to
align with the local magnetic field it experiences, like a little compass. In 
the
so-called ``Ising model,'' the spins are constrained to a given direction,
may point upward ($m({\bf r}) =+1$) or downward ($m({\bf r})
=+1$). The average magnetization is defined as 
\begin{equation}
M=\frac 1N\sum_{{\bf r}}m({\bf r}) , 
\end{equation}
where $N\rightarrow \infty $ is the total number of spins, and
the local field is composed of the applied
field and the one created by neighboring spins. On a mean-field level,
one assumes that the system may still be faithfully described if the many
degrees of freedom are reduced to a single quantity, here the average
magnetization. This assumption becomes exact if the interaction between spins 
is infinite
in range, that is, all spins interact with all others in the same
way, so that they contribute to the local field equally everywhere.
Even if the coupling has a limited range, we can imagine that such
an approach becomes more and more accurate as the dimensionality is
increased, as spins acquire more and more neighbors, thus
``averaging out'' their contributions to the local field. In fact, 
the mean-field approximation yields the correct critical
behavior above a given, {\it finite} dimension (see Sec.~V). 

\begin{figure} [h]
\epsfxsize=9truecm 
\centerline{
\epsfbox{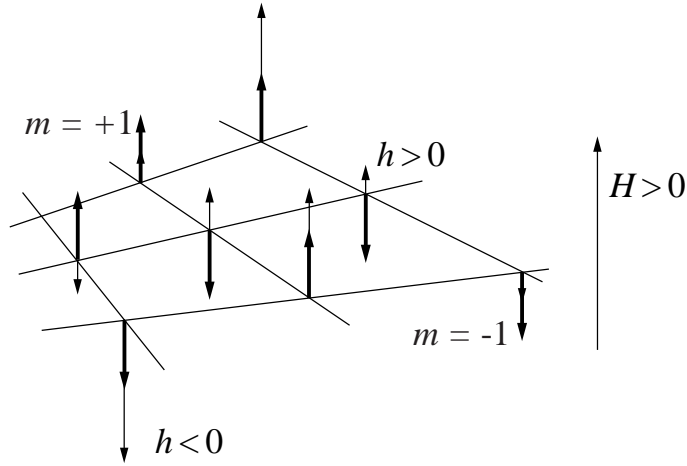} 
} 
\vspace*{.4truecm}

\caption{Illustration of a few ``sites'' of a two-dimensional Ising model with 
a random field. The thick arrows represent the spins which can point upward ($m=+1$) or downward ($m=-1$), and the thin arrows represent the random field
$h$. Each spin experiences a uniform applied field $H$.}
\vspace*{.5truecm}
\end{figure}

We confine our study to  zero temperature, so that spins are always
perfectly aligned with the local field, forbidding any {\it
thermal} fluctuation. In real materials, however, the presence of
impurities leads to {\it random} fluctuations. In many cases, the
effect of the impurities is to perturb the local magnetic field
which acquires a stochastic component in addition to a uniform one
(Fig.~9)\cite{rfim}. More precisely, the mean-field, driven, 
disordered version of the Ising model, first introduced in
Ref.~\ref{bib:sethna}, is defined by a local field
\begin{equation}
\label{local}f({\bf r}) =H+h({\bf r})
+JM\text{,} 
\end{equation}
where $H$ is a uniform field (chosen by the experimentalist), 
$h$ is a {\it random field}
chosen at each position ${\bf r}$ from some probability distribution 
$p\left( h\right)$, and $J>0$
represents the strength of the spin-spin coupling. The configuration of the 
system is then given
by the ``alignment rule'' 
\begin{equation}
\label{rule}\left\{ 
\begin{array}{c}
m({\bf r}) =-1\quad \text{if}\quad f({\bf r})
<0\Leftrightarrow h({\bf r}) <-H-JM, \\ m({\bf
r}) =+1\quad \text{if}\quad f({\bf r})
>0\Leftrightarrow h({\bf r}) >-H-JM. 
\end{array}
\right. 
\end{equation}

How does the driven disordered Ising model evolve when the
(driving) field $H$ goes from
$-\infty$ to $\infty$? Initially ($H=-\infty$), all the spins
point downward ($m=-1$). As $H$ is increased from $-\infty$,  the
local fields with large positive random fields soon become
positive, and the associated spins flip upward. These flips further
increases the local field, which induces new spin flips, and so on.
Once $H$ is incremented enough, the negative $h$'s tend to pin the
corresponding spins downward, and the configuration results from
the competition between the driving agent $H$, the ``restoring''
force $JM$, and  the pinning forces $h$ --- a common scenario for
forced or driven random media\cite{kardar,fisher}.

The avalanching present in this simple model is
reminiscent of fracture phenomena\cite{zapperi}, and the driven
disordered Ising model can be mapped to the \DFBM\ as 
$$
\begin{array}{ccc}
\text{downward spin (}m=-1\text{)} & \longleftrightarrow & \text{intact
fiber,} \\ \text{upward spin (}m=+1\text{)} & \longleftrightarrow &
\text{broken fiber,} \\ \text{random field (}h\text{)} &
\longleftrightarrow & 
\text{strength threshold,} \\ \text{uniform field (}H\text{)} & 
\longleftrightarrow & \text{driving force,} \\ \text{spin-spin coupling (}J 
\text{)} & \longleftrightarrow & \text{redistribution of the force.} 
\end{array}
$$
As in the case of the boarding gate problem, the main difference with the
\DFBM\ arises from the way in which the individual degrees of
freedom interact, and the iterative equation for the fraction
$n_{\downarrow}$ of downward spins, in the driven disordered Ising
model, reads 
\begin{equation}
\label{iteration1}n_{\downarrow i+1}=P( -H-JM_i) , 
\end{equation}
where $P(x) =\int_{-\infty }^xp(h)
dh$ is the cumulative distribution function. Equivalently, in
terms of the magnetization $M=1-2n_{\downarrow}$, 
\begin{equation}
\label{iteration2}M_{i+1}=1-2P(-H-JM_i) . 
\end{equation}
As before, the right-hand-side is a monotonically increasing function of the
avalanching variable $M_i$, and the so-called {\it hysteresis} curve $
M=M_\infty (H)$ is given by the left-most intersection of $y(
M) \equiv 1-2P(-H-JM)$ with the bisector ($y=x$).

Rather than considering all possible types of disorders, we focus on three,
commonly encountered in the literature\cite{schneider,aharony}.

\noindent(1) {\it Gaussian disorder}: $p(h( {\bf r})) =\frac
1{\sqrt{2\pi \sigma^2}}e^{-h^2/2\sigma^2}$. As $H$ is increased,
the curve $y(M)$ is translated to the left and the
intersection point consequently moves up (and to the right). Three
qualitatively different solutions are obtained, depending on
whether the maximum slope $y^{\prime}( M=-H/J)
=2J/\sqrt{2\pi \sigma^2}$ is greater, equal, or lesser than 1
(Fig.~10).

\begin{figure} 
\epsfxsize=9.6truecm 
\vspace*{.4truecm} 
\centerline{
\epsfbox{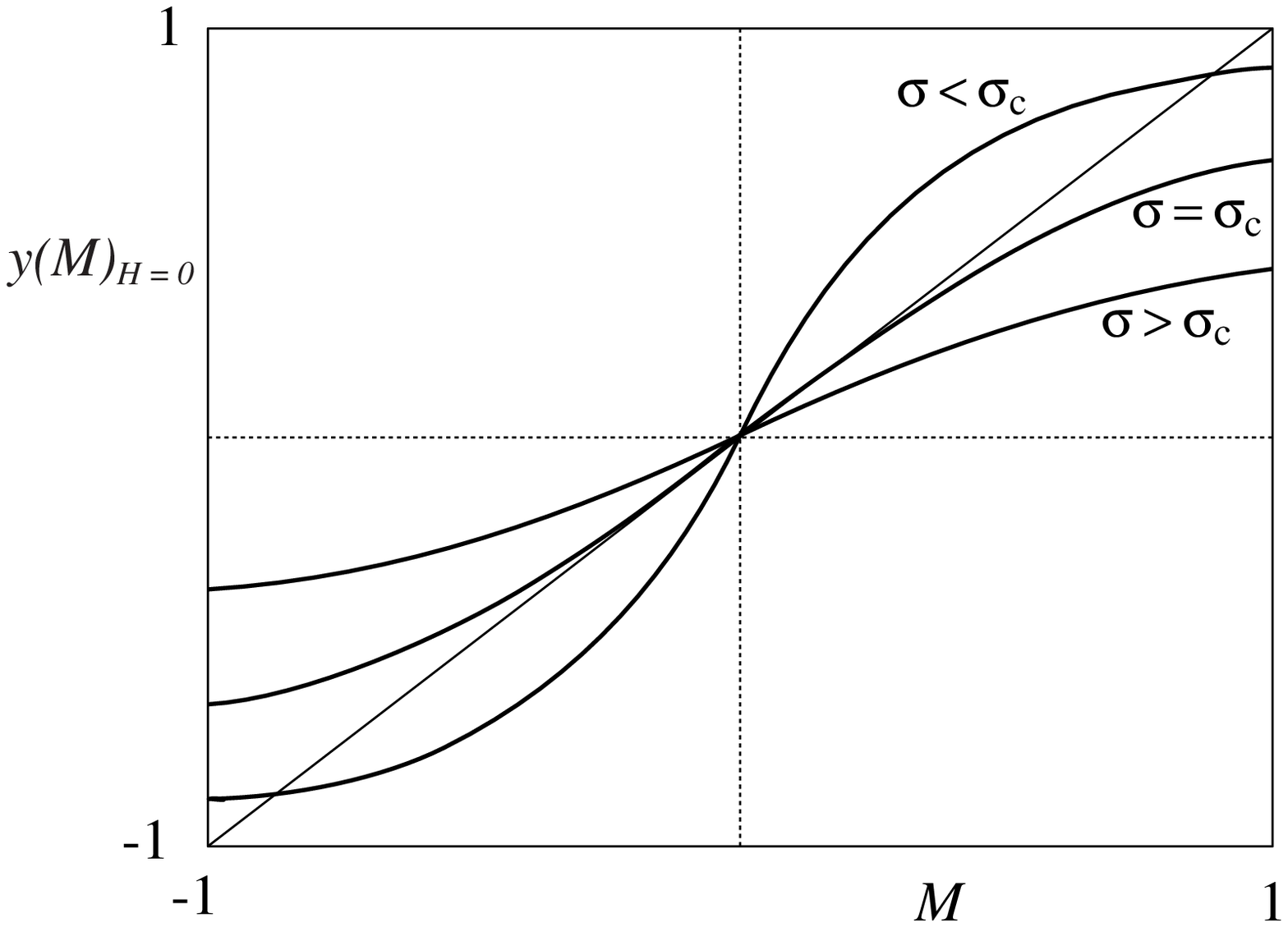} 
}
\epsfxsize=9truecm 
\vspace*{.7truecm} 
\centerline{
\epsfbox{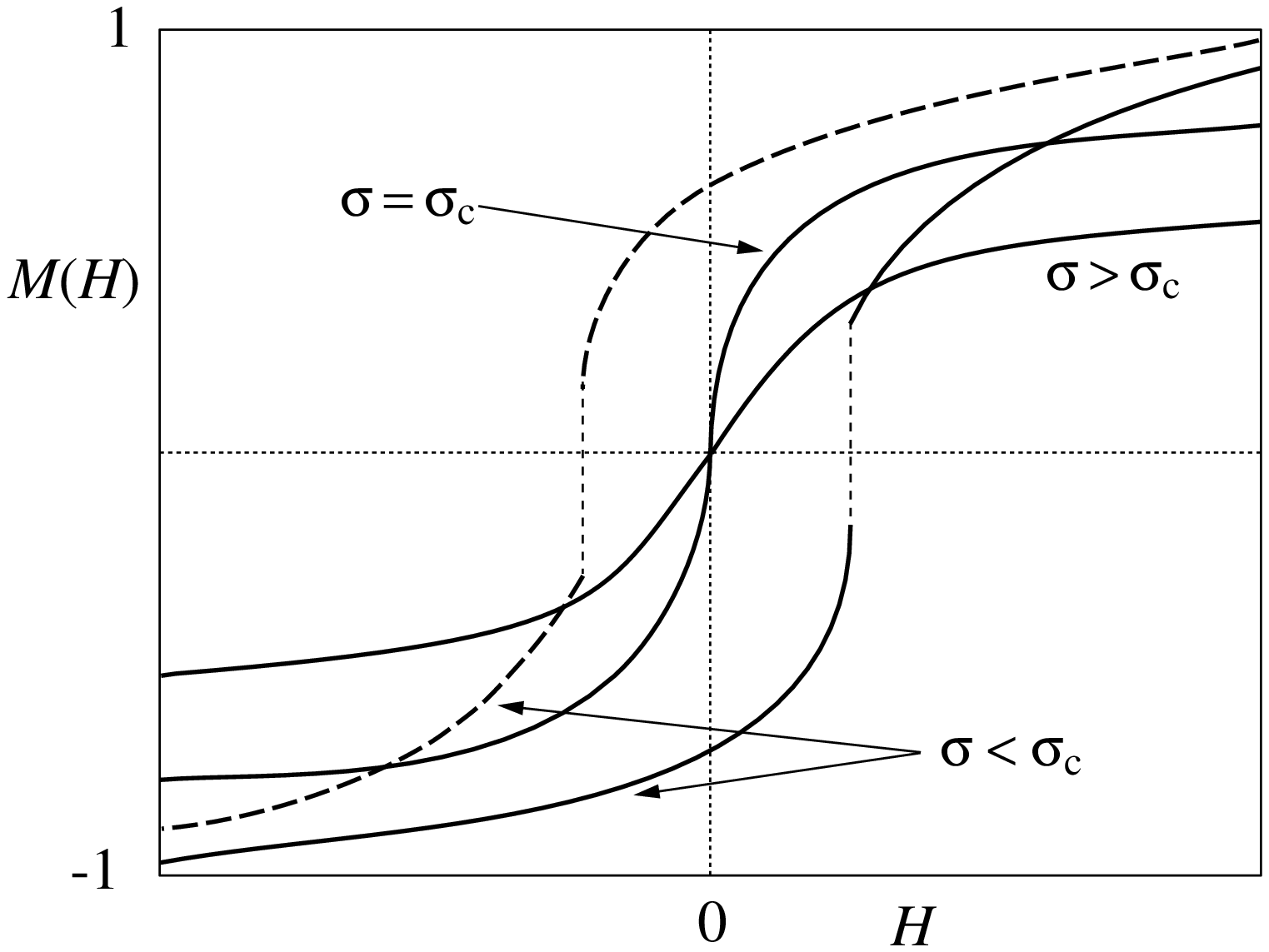} 
}  
\vspace*{.5truecm}

\caption{The driven Ising model with a Gaussian random field.
(a) Illustration of the graphical scheme for sub-critical, critical, and 
super-critical values of the disorder. The curves $y(M)$ are
drawn at $H=0$; as $H$ is increased, they move to the left,
starting in the far  right at large negative $H$ and ending in the
far left at large positive $H$. (b) Corresponding typical
magnetization, or hysteresis, curves. The dashed line represents
the case of a decreasing $H$, in the sub-critical regime. In the 
critical and super-critical regimes, the curves for increasing and 
decreasing $H$ are superposed.}
\vspace*{.6truecm}
\end{figure}

\noindent $\bullet$ $\sigma < \sigma_c \equiv \sqrt{2/\pi}J$: For a
small disorder, there is a jump in the magnetization at some field
$H_{\text{jump}}>0$), signaled, as in the \DFBM, by a square
root divergence of the response function or susceptibility
\begin{equation}
\label{divergence}\chi =\frac{dM_\infty }{dH}\sim (
H_{\text{jump}}-H)^{-1/2}. 
\end{equation}
Larger values of $\sigma$ reduce the magnitude of the
discontinuity, which ultimately shrinks to a point at $\sigma
=\sqrt{2/\pi }J$.

\noindent $\bullet$ $\sigma =\sigma_c$: A specific value
$\sigma_c$ of the disorder yields a {\it  critical}
hysteresis curve, which, although continuous, displays a singular
point at $H=M=0$. Expanding
$y(M)$, we easily obtain the nature of the
singularity, as 
\begin{equation}
\label{susceptibility}\chi(H\rightarrow 0) \sim H^{-2/3}. 
\end{equation}
In a more general line of reasoning (see Sec.~V), we note that
the asymmetry of $y$ about the origin (at $H=0$) rules out even
powers relating $M$ to $H$ as 
\begin{equation}
H\sim M^{2n}+\text{ higher powers,} 
\end{equation}
and in particular a quadratic dependence $H\sim M^2$, leaving a cubic
behavior $H\sim M^3$ as the best candidate; inverting and differentiating
readily leads to Eq.~(\ref{susceptibility}).

\noindent $\bullet $ $\sigma >\sigma_c$: At large
disorders, the magnetization curve is smooth, with a finite susceptibility
everywhere.

Figure~10(b) shows generic
curves corresponding to the above three cases. This phase diagram is very
similar to that of the \DFBM\ (classes ({\it i})--({\it iii})), in
that a discontinuous, first-order-like region terminates at a
critical, or  second-order, point with a different (non-trivial)
exponent. In fact, it is believed\cite{dahmen1,dahmen2} that,
while the global phase diagram is unchanged in a local theory
extending beyond a mean-field approach, the  precursor
divergence of Eq.~(\ref{divergence}) is suppressed, resulting in proper
first-order behavior. (If a similar conclusion applies to fracture
phenomena, it might help to explain the presumably abrupt failures often
observed.) Furthermore, as $\sigma $ approaches $\sigma_c$ from
below, $H_{\text{jump}}$ decreases to zero, progressively hampering
the hysteretic (lagging) nature of the solution; for $\sigma \geq
\sigma_c$, both the remnant and coercive fields are vanishing.
Again, this is an artifact\cite{dahmen2} of the present
formulation.  These remarks point at the tip of an iceberg onto
which the mean-field approximation often runs.

\noindent(2) {\it Uniform disorder}: $p(h({\bf r})) =1/2\sigma$ for
$-\sigma \leq h \leq \sigma$ {\it and vanishes elsewhere}. This
form may be viewed as a ``digitized,'' roughened, or sharpened
version of Gaussian disorder. Again, the character of the solution
depends on the magnitude of the slope $y^{\prime }(
M=-H/J) =J/\sigma$ relative to $1$ (Fig.~11).

\begin{figure} 
\epsfxsize=9.6truecm 
\vspace*{.4truecm} 
\centerline{
\epsfbox{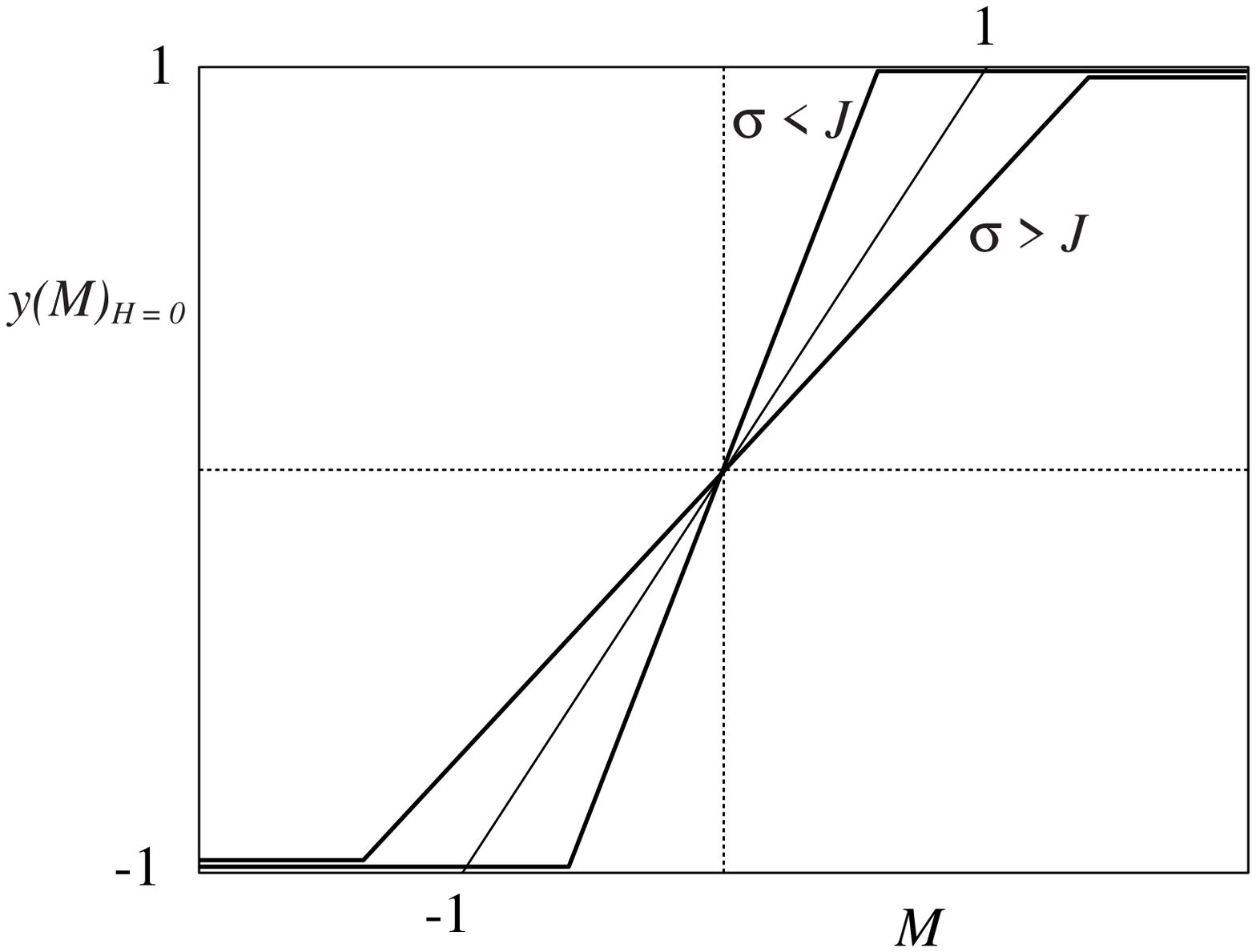} 
}
\end{figure}

\begin{figure}
\epsfxsize=9.2truecm 
\centerline{
\epsfbox{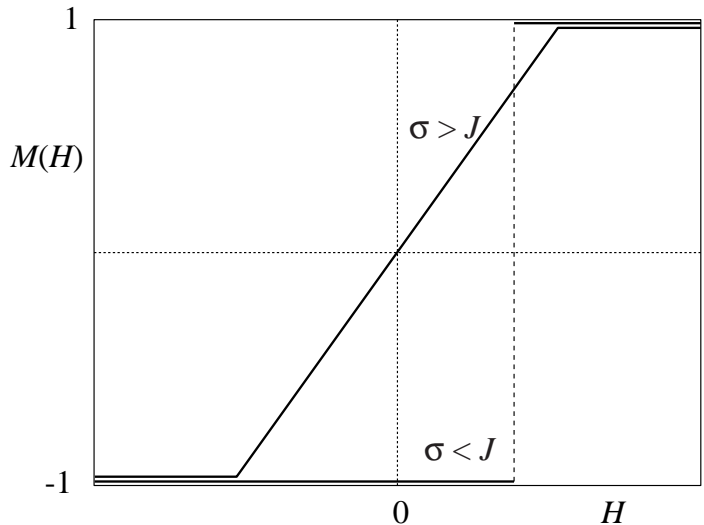} 
}  
\vspace*{.2truecm}

\caption{The driven Ising model with a uniform random field. (a) Illustration 
of the graphical scheme and (b) typical solutions.}
\vspace*{.1truecm}
\end{figure}

\noindent $\bullet$ If $\sigma <J$ (small disorder), the
magnetization switches from $-1$ to $+1$ at
$H_{\text{jump}}=J-\sigma >0$.

\noindent $\bullet$ If $\sigma >J$ (large disorder), the
magnetization is continuous and piecewise linear; it starts to
increase linearly from $-1$ at 
$H=J-\sigma $, and reaches $+1$ at $H=\sigma -J$.

As in the previous examples, the nature of the solution varies
significantly from one type of disorder to the next. In particular, we note
that the discontinuity of the uniform distribution rules out any divergence or
critical point, in agreement with our observations for the \DFBM.

\noindent(3) {\it Bimodal disorder}: $p(h( {\bf r})) =\left[ \delta
\left( h+\sigma \right) +\delta(h-\sigma)
\right]/2$. The bimodal distribution is often used as an
alternative to the
Gaussian distribution, notably in computer simulations. It represents an even
more ``digitized,'' or sharpened, version of disorder, reminiscent
of the
\DFBM\
with fibers of two different types. The graphical scheme is
illustrated in Fig.~12, along with typical solutions which fall, as
above, into one of two classes. The phase diagram is especially simple to
understand for a bimodal disorder, which limits random
fluctuations as much as possible and reduces the problem to a two-body one.
Half of the spins (those with $h=+\sigma $) flip upward when $H$ reaches $H_{
\text{jump1}}=J-\sigma $, leading to a vanishing magnetization. This further
triggers the flipping of the remaining spins if their (updated) local field
$f=H_{\text{jump1}}-\sigma $ is positive, that is, if $\sigma <J/2$
(small disorder), resulting in a global switch from $M=-1$ to
$M=+1$ at
$H_{\text{jump1}}$. If $\sigma >J/2$ (large disorder), on the
other hand, the magnetization remains at zero until the local field
changes sign, at $H_{
\text{jump2}}=\sigma $, when the remaining downward spins flip to yield
$M=+1$.

\begin{figure} 
\epsfxsize=9.7truecm 
\centerline{
\epsfbox{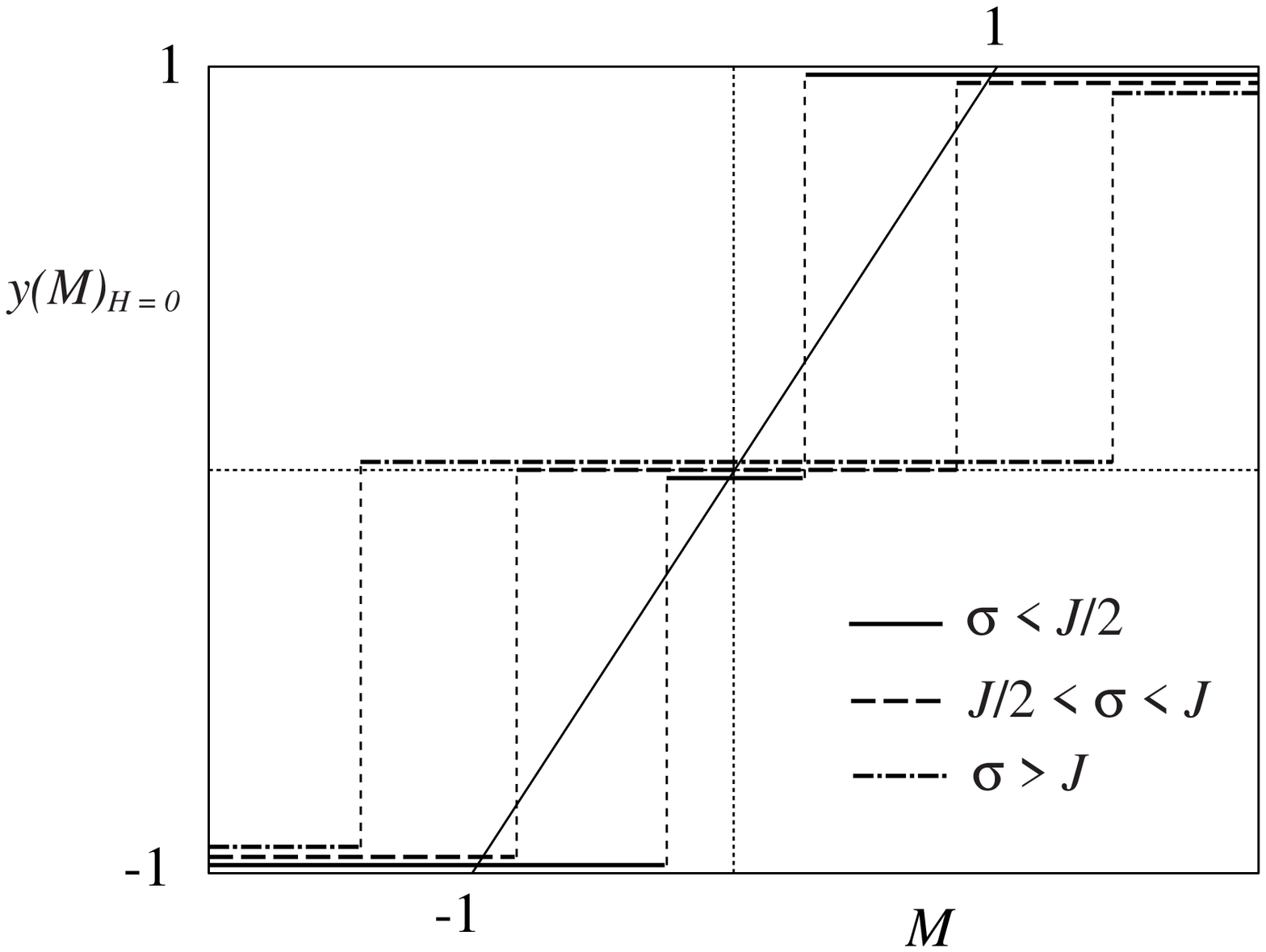} 
}
\end{figure}

\begin{figure}
\epsfxsize=9.2truecm 
\centerline{
\epsfbox{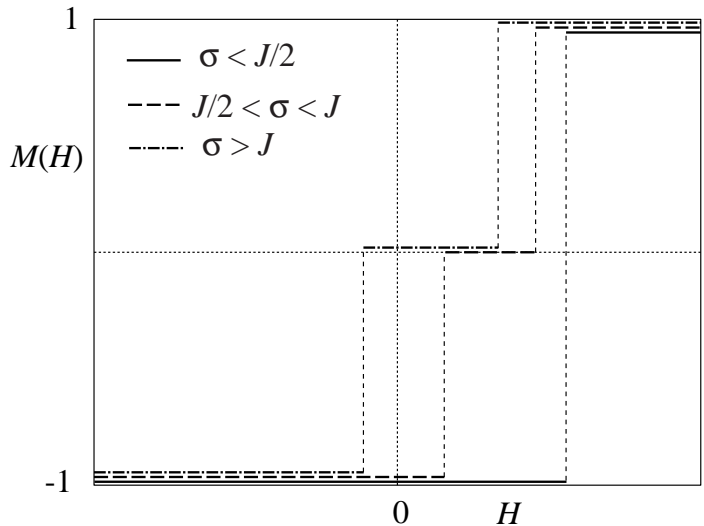} 
}  
\vspace*{.4truecm}

\caption{The driven Ising model with a bimodal random field. (a) Illustration 
of the graphical scheme and (b) typical solutions.}
\vspace*{.6truecm}
\end{figure}

In contrast to the Gaussian and uniform cases, the large-disorder
phase {\it exhibits hysteresis}: $M(H)$ is not symmetric
about $H=0$, there is a remnant magnetization $M=-1$ (if $H_{\text{jump1}}>0$
that is, $\sigma <J$) and a coercive field which can be defined as
$H_{
\text{jump2}}$. Clearly, a hysteretic behavior originates in the coexistence
of more than one possible solution to a dynamical
system\cite{silveira}.  Rather than being
chosen by an optimum principle, the actual, physical solution follows from
the {\it history}, or, in the case of a monotonically increasing
$H$, from the {\it initial conditions}. Thus, {\it metastability}
and hysteresis are intimately related. The random fluctuations tend
to pin the system to a metastable state. The thermal fluctuations,
which we have ignored, play an opposite part in sweeping the system
in its configuration space, thus favoring ergodicity. An important
question relates to the interplay of these two effects. It is
believed that disorder generally dominates and that thermal
fluctuations are irrelevant at criticality, both in
equilibrium\cite{rfim} and in
non-equilibrium\cite{sethna,dahmen1,dahmen2,silveira} problems.

\section{Digression: Effective Theories}

The path from elementary laws to the multitude of fascinating 
observed phenomena is definitely not a straight and plain one. The
situation is a bit like that of a shadow show in which shadows of
dragons, wolves, and birds, are cast on a white drape by the moving
hands of a pantomime. The goal is to understand the unfolding
story. Focusing on the artist's hands is probably not the most
fruitful tactic, and surely not the easiest. The shadows allow a
more direct grasp of the story by retaining the important aspects
of the hands' intricate movements and shapes, while smearing away
irrelevant details such as the nails or the hair on the fingers. In
theoretical physics, the mime's hands are the fundamental laws, and
the shadows are {\it effective} or {\it phenomenological} theories.

Hydrodynamics constitutes a prominent example of an effective theory,
where the molecular nature of a fluid is disregarded in favor of a
description in terms of a homogeneous medium.
The collective motion of hundreds or millions of particles
is reduced to that of their center of
mass, that is, to a local average, which
smears out a large number of details and fine characteristics,
leaving a framework with a smaller number of variables than
the original one. Typically, a fluid is then described by fields of
density, velocity vector, and temperature --- five quantities
associated with each point in space and time substitute for
the many molecular degrees of freedom around that point.

A delicate question
relates to how much of the finer print, or {\it fluctuations,} needs to be
included for an accurate description of a given phenomenon at
a given length scale. 
Various effective theories may be constructed to describe a given system, 
the ones with the fewest allowed
fluctuations being termed ``mean-field.'' Often, spatial
fluctuations are ignored altogether resulting in highly coherent behavior,
and the system reduces to a single (or a few) degrees of freedom. 

The hydrodynamic equations can be derived from symmetry and
conservation principles alone.
These principles are satisfied at the molecular level and unspoiled
by any averaging procedure. Thus, an effective theory ensues from
robust properties that span many length scales, and is free from a
specific (mechanistic) model. It has lost the details involved in
the latter, but gained the sturdiness of generality: an effective
formulation encompasses all possible models lying within the
constraints imposed by a set of symmetries.

It is therefore worthwhile to construct a
phenomenological theory of breakdown phenomena; in what follows, we shall
concentrate on the driven disordered Ising
model\cite{sethna,dahmen2}. In addition to its robustness,  an
effective field theory offers an analytical framework in which the
range of the coupling between spins, the dimensionality of the
systems, or other aspects may be modified and studied in a unified 
fashion, and for which a number of powerful tools have been
developed\cite{landau}.

In the spirit of hydrodynamics, we consider a spin {\it field} or {\it 
density} $s({\bf r})$ regarded as a local average of
magnetic moments at ${\bf r}\in \Re^d$, where $d$ is the dimension
of the system. Its value lies anywhere within an interval centered
at zero, and we further generalize the problem by allowing $s({\bf
r})
\in (-\infty ,\infty)$. Inspired by the model\cite{sethna,dahmen2}
of Sec.~IV and its solution as given by $y(M)-M=0$, we posit that
the spin density at each point is obtained from an algebraic
equation 
\begin{equation}
\label{motion}f_{{\bf r}}(\{s( {\bf r})\}
,H,h( {\bf r})) =0\quad \text{(for all }{\bf
r}\text{),} 
\end{equation}
where $H$ is a uniform applied magnetic field and $h({\bf
r})$ is a Gaussian random field felt by the spins at ${\bf
r}$. This system of equations may have more than one solution; if
$H$ is monotonically increased as before, the spins start off very
negative, and the physical solution corresponds to the
``left-most'' one, with smallest (most negative) $s({\bf
r})$'s. If, rather, $H$ is monotonically decreased from a
high positive value, we expect the same set of solutions with an
opposite sign, consistent with the {\it reflection symmetry} 
\begin{equation}
f_{{\bf r}}(\{ -s( {\bf r})\} ,-H,-h(
{\bf  r})) =f_{{\bf r}}(\{ s({\bf
r})\} ,H,h( {\bf r})) , 
\end{equation}
a first constraint on the form of $f_{{\bf r}}$. As before, the quantities of 
interest are
the overall spatial average $\left\langle s\right\rangle$ of
$s({\bf r})$ and its response to the applied field. An
effective, coarse-grained theory is clearly best suited to answer
questions on the system's global properties, and in particular on
its critical behavior  which involves a collective response to an
infinitesimal change in $H$. From the study of the driven
disordered Ising model, we expect the critical point to occur at
$H=\left\langle s\right\rangle =0$. (We note that this is also a
special point in terms of the symmetries, due to the {\it
statistical invariance} under
$h\rightarrow -h$.) Close to the latter, we may describe $f_{{\bf
r}}$ by an expansion in its arguments. In contrast to fluid
dynamics or pure critical phenomena, here quenched disorder imposes
some spatial fluctuations, which we limit by considering a small
random field, leading to

\begin{eqnarray} \label{general}
f_{{\bf r}} & = & \! \int \! d^dr^{\prime}J( 
{\bf r},{\bf r}^{\prime }) s( {\bf r}^{\prime})
+AH+Bh({\bf r}) \nonumber \\  &  & + \! \int \! d^dr^{\prime
}d^dr^{\prime
\prime}d^dr^{\prime \prime \prime}K({\bf r},{\bf r}^{\prime},{\bf
r}^{\prime \prime},{\bf r}^{\prime
\prime
\prime})\, s({\bf r}^{\prime}) s({\bf r}^{\prime
\prime}) s({\bf r}^{\prime \prime \prime})
\nonumber \\  &
&
+\,H \! \int \! d^dr^{\prime}d^dr^{\prime \prime}L( {\bf r},{\bf
r}^{\prime},{\bf r}^{\prime \prime})\, s({\bf r}^{\prime}) s( {\bf
r}^{\prime \prime}) +h({\bf r}) \! \int \! d^dr^{\prime}d^dr^{\prime
\prime}M({\bf r},{\bf r}^{\prime},{\bf r}^{\prime \prime})\,
s({\bf r}^{\prime} s({\bf r}^{\prime \prime})
\nonumber \\  &  & +\, H^2 \! \int \! d^dr^{\prime}N( 
{\bf r},{\bf r}^{\prime})\, s( {\bf r}^{\prime})
+h({\bf r})^2 \! \int \! d^dr^{\prime}O( {\bf r},{\bf
r}^{\prime })\, s({\bf r}^{\prime }) +H h( 
{\bf r})\! \int \! d^dr^{\prime}P( {\bf r},{\bf r}^{\prime
})\, s( {\bf r}^{\prime }) \nonumber \\  &  &
+\,CH^3+Dh({\bf r})^3 \nonumber +\ldots
\end{eqnarray}
Along the lines of Eq.~(\ref{general}), where we have written
the most general expansion consistent with the
governing symmetries, we further expand the interactions.
For example, we write
\begin{eqnarray}
\int \! d^dr^{\prime}J( {\bf r},{\bf r}^{\prime })
s({\bf r}^{\prime}) & = & \! \int \! d^dr^{\prime }J( 
{\bf r},{\bf r}^{\prime }) \left\{ s( {\bf r})
+\sum_\alpha
\left(r_\alpha -r_\alpha^{\prime }\right) \frac \partial {\partial
r_\alpha }s( {\bf r}) \right. \nonumber \\  &  & \left.
+\frac 1{2!}\sum_{\alpha ,\beta }\left( r_\alpha -r_\alpha^{\prime
}\right) \left(r_\beta -r_\beta^{\prime }\right) 
\frac{\partial^2}{\partial r_\beta \partial r_\alpha }s({\bf
r}) +\ldots \right\} \nonumber \\  & = & \left\{ \! \int \!
d^dr^{\prime  }J(
{\bf r},{\bf r}^{\prime }) \right\} s( {\bf r})
+\sum_\alpha \left\{ \! \int \! d^dr^{\prime}\left( r_\alpha
-r_\alpha^{\prime }\right) J( {\bf r},{\bf r}^{\prime})
\right\} \frac \partial {\partial r_\alpha }s({\bf r})
\nonumber \\  &  & +\frac  1{2!}\sum_{\alpha
,\beta }\left\{ \! \int \! d^dr^{\prime}\left( r_\alpha
-r_\alpha^{\prime }\right) \left( r_\beta -r_\beta^{\prime }\right)
J({\bf r},{\bf r}^{\prime }) \right\}
\frac{\partial^2}{\partial r_\beta \partial r_\alpha }s({\bf
r}) +\ldots
\label{expansion}
\end{eqnarray}
If the couplings are invariant under translations and rotations, then
$J({\bf r},{\bf r}^{\prime }) =J( \left| {\bf r}-{\bf
r}^{\prime }\right|)$, the odd powers in
Eq.~(\ref{expansion}) vanish and the even powers take a symmetric
form, resulting in 
\begin{equation}
\int \! d^dr^{\prime }J( {\bf r},{\bf r}^{\prime })
s( {\bf r}^{\prime }) =as( {\bf r})
+b\nabla^2s( {\bf r}) +\ldots , 
\end{equation}
where $a$ and $b$ are numerical coefficients obtained from $J$. The
derivative terms couple neighboring spins by hampering (or favoring,
depending on the sign of the prefactors) spatial fluctuations. In their
interplay with disorder effects, such short-range interactions introduce
higher complications. By analogy
with the driven disordered Ising model, we consider a mean-field
approximation by substituting the derivative interactions
with a uniform restoring force composed of $\left\langle s\right\rangle $, $
\left\langle s\right\rangle^3$, $H\left\langle s\right\rangle^2$,
$H^2\left\langle s\right\rangle$, etc. If a similar procedure
is applied to the various terms in $f_{{\bf r}}$, we obtain 
\begin{eqnarray}
f_{{\bf r}}&=&as( {\bf r}) +b^{\prime}\left\langle
s\right\rangle +AH+Bh({\bf r}) +cs({\bf
r})^3 \nonumber \\ & & +\left( \text{other
terms, such as }\left\langle s\right\rangle^3 \text{, }Hs^2\text{, }h^2s\text{
, or }H^3\right) . 
\end{eqnarray}
To get a flavor of the construction and treatment of a
phenomenological theory, rather than to examine the problem in full rigor, we
abandon the ``other terms'' for now --- we argue below that they do not
change the critical properties. Furthermore, for ease of notation, we
shall not drag along the many prefactors; the reader may either imagine them
present or assume that they have been set to unity. Thus, our task is now to
solve Eq.~(\ref{motion}), which has been reduced to
\begin{equation}
\label{simple}s( {\bf r}) +\left\langle s\right\rangle
-H+h({\bf r}) +s({\bf r})^3=0. 
\end{equation}
The prefactor of $H$ has been set to $-1$ so that $s$ aligns
along $H$ at large values of the latter. This {\it stability condition} 
is imposed in addition to the
various symmetries. In solving Eq.~(\ref{simple}), we think of
$\left\langle s\right\rangle$ as a
given function of $H$. If the
resulting $s({\bf r})$ does not average to $\left\langle
s\right\rangle$, our choice was not good, and we start over with a
better guess for $\left\langle s\right\rangle$, until eventually the
spatial average of $s({\bf r})$ is equal to
$\left\langle s\right\rangle$ for all $H$. In practice,
Eq.~(\ref{simple}) may be solved  iteratively
by reshuffling the terms, as
\begin{eqnarray}
s({\bf r}) & = & -\left\langle s\right\rangle +H-h( 
{\bf r}) -s({\bf r})^3 \nonumber \\  & = &
-\left\langle s\right\rangle +H-h({\bf r}) -\left[
-\left\langle s\right\rangle +H-h({\bf r})
\right]^3-\ldots 
\label{iter_sol} \end{eqnarray}
Clearly, performing the average over space or over disorder is
equivalent, and the self-consistency condition results in 
\begin{equation}
\label{self-cons}(3\sigma^2-2) \left\langle
s\right\rangle +\left\langle s\right\rangle^3=(
3\sigma^2-1) H +\left(\text{higher-order terms}\right), 
\end{equation}
where $\sigma^2=\left\langle h^2\right\rangle$. For large
disorders, the magnetization $\left\langle s\right\rangle$ is a
non-singular function of $H$. At a critical value $\sigma^2=2/3
\equiv
\sigma_c^2$, however, 
$\left\langle s\right\rangle \propto H^{1/3}$ and the response
function diverges as 
\begin{equation}
\chi =\left. \frac{\partial \left\langle s\right\rangle }{\partial
H}\right|_{H \rightarrow 0}\sim H^{-2/3}, 
\end{equation}
reproducing the critical exponent of the driven disordered Ising
model.

Let us return to the ``other terms'' neglected in Eq.~(\ref{simple}),
and argue that they do not modify the exponents. Indeed, any one of them
either vanishes upon averaging ($h^3$ for example) or is subleading in the
limit $H,\left\langle s\right\rangle \rightarrow 0$ (for example, $H^2s$
yields $H^3$ and $H^2\left\langle s\right\rangle $ upon averaging). In
a third possibility, the additional terms modify the prefactors 
in Eq.~(\ref{self-cons}) (for example,
$h^2s$ yields $\sigma^2\left\langle s\right\rangle$ upon
averaging), but do not generate new types of terms. The symmetries are
constraining enough to impose the cubic root behavior of the magnetization
close to criticality. In a mean-field approach, in particular, where the
system is reduced to a single degree of freedom, the exponents follow from
an algebraic equation such as Eq.~(\ref{self-cons}). This observation sheds 
some light on
the (then quite cryptic) remark in Sec.~II, that an exponent $-1/2$ ``is to be
expected within a mean-field approach.'' Without the reflection
symmetry that we have assumed in the derivation of
Eq.~(\ref{self-cons}), quadratic terms would appear and
clearly lead to an inverse square root divergence in the response function.

Another way to understand the decoupling that occurs in a mean-field
theory and its essentially zero-dimensional nature (single degree of
freedom) is obtained
in a formulation closer to equilibrium statistical mechanics where the
central objects are probability weights and partition functions.
{}From rewriting Eq.~(\ref{simple}) as 
\begin{equation}
-h( {\bf r}) =s( {\bf r}) +\left\langle
s\right\rangle -H+s( {\bf r})^3 
\end{equation}
and the random field density 
\begin{equation}
p(h) ={\cal N}e^{-\frac{h^2}{2\sigma^2}}, 
\end{equation}
where ${\cal N}$ is a normalizing factor, the spin field is distributed
according to 
\begin{equation}
\label{weight}\rho \left( s({\bf r}) \right) ={\cal N}\exp
\left[-\frac 1{2\sigma^2}\left( s({\bf r})
+\left\langle s\right\rangle -H+s({\bf
r})^3\right)^2\right] \times \left| 
\left|
\text{Jacobian}\right| \right| . 
\end{equation}
Again, we require the self-consistency condition
\begin{equation} \label{self}
\left\langle s\right\rangle =\int ds\;s\;\rho(s) , 
\end{equation}
and the problem reduces to a zero dimensional one: finding
the average of $s$ at a single ``site.''
The right-hand-side of Eq.~(\ref{self}) may be 
evaluated by expanding the exponential in the super-quadratic
terms of its argument and calculating Gaussian integrals.

In the absence of modulated interactions,
the Jacobian is easily evaluated as $1+3s$. Beyond the mean-field
picture, however, fluctuations matter regardless of their coupling to $h$,
and the need for and use of a Jacobian becomes quite cumbersome. For a
brief (and very superficial) discussion of richer theories, we return to the
formulation of Eq.~(\ref{simple}), whose analog for the case of short-range
interactions reads 
\begin{equation}
\label{short}s({\bf r}) +\nabla^2s({\bf r})
-H+h({\bf r}) +s({\bf r})^3=0. 
\end{equation}
How does the coupling between neighboring spins,
expressed in the Laplacian term, modify the critical behavior?
This question is best examined through the
renormalization-group theory\cite{rg}, which provides a systematic
way to keep track of the role of each intervening term. Roughly
speaking, it enables us to carry out an iterative solution (or
averaging), similar to that of Eq.~(\ref{iter_sol}), in successive
steps corresponding to progressively larger scales.
Interestingly, it is found that the answer depends only on the 
dimension $d$ of the system.
Above six dimensions, a system described by Eq.~(\ref{short}) behaves
just as the mean-field one of
Eq.~(\ref{simple})\cite{dahmen1,dahmen2}, as  far as critical
exponents go. In some sense, it is bundled tightly enough so that a
small push might move a lot. (Many hay stems are often dragged
along when one is pulled out of a three-dimensional pile; this does
not happen if the hay stems are laying on the two
dimensional surface of a pond.) In $ d<6 \equiv d_c$
(called ``upper critical dimension''), the situation is far less
trivial, and novel exponents emerge from the interplay of the
Laplacian and random terms\cite{dahmen1,dahmen2}. We might  also
expect intuitively that longer ranged interactions ``more easily''
lead to mean-field behavior and, indeed, they reduce the upper
critical dimension. The case in which $d_c<3$ explains the trivial,
mean-field-like exponents associated with some observed critical
phenomena  that, {\it a priori}, involve many complicated
interactions.

In some instances, fluctuations bring about further surprises. A
mean-field approximation somehow ``homogenizes'' the system,
imposing a high degree of symmetry. Fluctuations might then break
the latter into lower symmetries: a flat surface is invariant under
any translation, whereas a periodically corrugated one is unchanged
only under translations by the period's multiples. New critical
points then appear, corresponding to the new symmetries. This is
exactly what happens in a system defined by the vectorial
generalization of Eq.~(\ref{simple})\cite{silveira}, with the new
degrees of freedom $\vec s({\bf r}) \in \Re^n$. At the
mean-field level, the critical properties correspond to a full
rotational symmetry (in spin space). For short-range interactions,
new critical points appear, associated with rotational invariance
in subspaces of $\Re^n$\cite{silveira}.

It is truly remarkable that we are able to make precise predictions
based on such a general approach, without a
definite model or ``fundamental'' understanding of the details. It
is equally surprising that a complicated assembly of electrons and
nuclei generates pure numbers, the critical exponents, which can be
measured with great accuracy.

\section{Conclusion}

The models examined in Secs~II--IV are probably less
interesting for their own sake than for their illustrative power.
Also, they provide motivation and inspiration for more complex,
robust formulations closer to physical reality. This is especially
true of simple models involving ubiquitous features such as driving
forces, elastic interactions, and disorder. The breakdown of a
disordered conductor under an applied current or voltage drop
(often thought of as a jumping board to fracture problems) is yet
another illustration. In a simple formulation\cite{zapperi}, it
may in fact be mapped to the driven disordered Ising model and the
graphical scheme presented here is readily applicable to it. 

The ability to solve such basic models serves as a first
step toward the study of richer, more realistic physics. A class of problems 
akin
to breakdown phenomena relates to the depinning and transport of elastic
manifolds through disordered media\cite{kardar,fisher}. The many
variations  on this paradigmatic
theme include charge density waves\cite{fisher}, surface
growth\cite{barabasi,kardar}, contact
lines\cite{de_gennes,ertas,fisher},  polymers\cite{kardar}, flux
lines in superconductors\cite{kardar,fisher}, 
fractures\cite{fisher}, and earthquakes\cite{fisher}. Another
class of related examples pertains to social  dilemmas, as
illustrated by financial markets\cite{mandelbrot,bouchaud}, traffic
flow, and internet congestion problems\cite{huberman}. Given this
wide range of interesting systems, it is worthwhile to acquire an
elementary understanding and intuition from some simple related
models. A rough ``conceptual web'' involved in the three examples
we discussed in some detail is illustrated in the following diagram.

$$
\begin{array}{ccc}
\text{driving force}
& 
\begin{array}{c}
\text{coupling} \\ \text{(restoring force)}
\end{array} 
& 
\begin{array}{c}
\text{disorder} \\ \text{(pinning force)}
\end{array} 
\\ \searrow  & \downarrow  & \swarrow  \\  
& \quad \text{breakdown phenomenon} & 
\end{array}
$$
$$
\underbrace{
\begin{array}{cc} 
/ &  \backslash  \\ 
\text{metastability}
&
\quad 
\begin{array}{c}
\text{(non-equilibrium)} \\ \text{critical phenomena}
\end{array} 
\end{array}
} 
$$
$$
\underbrace{
\begin{array}{ccc}
\text{symmetry} & \text{and} & \text{universality classes} \\ \downarrow  &  
& \downarrow  \\ 
\text{hysteresis} &  & 
\begin{array}{c}
\text{relevance of the} \\ \text{type of disorder}
\end{array}
\end{array}
} 
$$
$$
\begin{array}{c}
\uparrow  \\ 
\text{graphical scheme}
\end{array}
$$
\\

The main gap between our discussion and a more complete description of
reality is the absence of any {\it dynamics}\cite{dynamics} in our 
formulation. It
takes some time for an elevator rope to break, during which stress is
redistributed. At a boarding gate, sitting people are watching,
thinking, and deciding {\it while} an impatient passenger walks up to the
line. A driven magnet responds with some delay, and different parts of the
system are continually changing while they influence one another. By
contrast, our formulation proposes algebraic equations to describe the
system rather than equations of motion. In the spirit of effective theories,
we can extend Eqs.~(\ref{simple}) and (\ref{short}) to include time 
dependences by
letting $s({\bf r})$ depend explicitly on time and replacing
the vanishing right-hand-side by a series such as 
\begin{equation}
\alpha \frac \partial {\partial t}s({\bf r},t) +\beta \frac{
\partial^2}{\partial t^2}s({\bf r},t) +\ldots
\end{equation}
This modification introduces correlations in time
and leads to a wealth of results relating to
fluctuations in time\cite{dynamics}.

\section*{Acknowledgments}

It is a pleasure to acknowledge Professor Mehran Kardar's precious
guidance and many interesting discussions. The author is
also grateful to Paul de Sa, as well as to Sohrab Ismail-Beigi and
Joel Moore, for commenting on the manuscript, and to the editors of
the theme issue, Harvey Gould and Jan Tobochnik, for their careful
review. This work was supported by the NSF through Grant No.
DMR-98-05833.


\begin{thebibliography}{99}

\bibitem{chakrabarti}For a general introduction and review on conductor 
breakdown, fractures, and earthquakes, see B. K. Chakrabarti and L. G. 
Benguigui,
{\sl Statistical Physics of Fracture and Breakdown in Disordered Systems} 
(Clarendon Press, Oxford, 1997).

\bibitem{peirce} F. T. Peirce, ``Tensile tests for cotton yarns. V.
-- `The weakest link', Theorems on the strength of long and
composite specimens,'' J. Textile Inst. {\bf 17}, T355--368 (1926).

\bibitem{daniels} H. E. Daniels, ``The statistical theory of the
strength of bundles of threads. I,'' Proc. R. Soc. London A {\bf
183}, 405--435 (1945).

\bibitem{comment} R. da Silveira, ``Comment on `tricritical behavior
in rupture induced by disorder','' Phys. Rev. Lett. {\bf 80}, 3157
(1998).

\bibitem{zapperi} S. Zapperi, Purusattam Ray, H. E.
Stanley, and A. Vespignani, ``First-order transition in the
breakdown of disordered media,'' Phys. Rev. Lett. {\bf 78},
1408--1411 (1997).

\bibitem{sethna} J. P. Sethna, K. Dahmen, S. Kartha, J. A. 
Krumhansl, B. W. Roberts, and J. D. Shore, ``Hysteresis and
hierarchies: dynamics of disorder-driven first-order phase
transformations,'' Phys. Rev. Lett. {\bf 70}, 3347--3350
(1993).~\label{bib:sethna}

\bibitem{andersen} J. V. Andersen, D. Sornette, and K.-t Leung,
``Tricritical behavior in rupture induced by disorder,'' Phys. Rev.
Lett. {\bf 78}, 2140--2143 (1997).

\bibitem{sornette} D. Sornette, ``Mean-field solution of a
block-spring model of earthquakes,'' J. Phys. I France {\bf 2},
2089--2096 (1992).

\bibitem{wu} B. Q. Wu and P. L. Leath, ``Failure probabilities and
tough-brittle crossover of heterogeneous materials with continuous
disorder,'' preprint, cond-mat/9811044 and references therein.

\bibitem{garciamartin} A. Garciamart\'{\i}n, A. Guarino, L. Bellon, 
and S. Ciliberto,
``Statistical properties of fracture precursors,'' Phys. Rev.
Lett. {\bf 79}, 3202--3205 (1997).

\bibitem{ising} E. Ising, Z. Physik {\bf 31}, 253 (1925). For an
introductory review, see Ref.~\ref{bib:huang}, Chapters 14 and 15.
For a more advanced treatment, see Ref.~\ref{bib:parisi}, Chapters
3 and 4.

\bibitem{huang}K. Huang, {\sl Statistical Mechanics}
(John Wiley and Sons, New York, 1987).~\label{bib:huang}

\bibitem{parisi}G. Parisi,
{\sl Statistical Field Theory} (Addison-Wesley Publishing Company,
Redwood City, CA, 1988).~\label{bib:parisi}

\bibitem{rfim} For a review of the random-field Ising model, see T.
Nattermann, ``Theory of the random-field Ising model,'' in {\sl Spin
Glasses and Random Fields}, ed. A. P. Young (World Scientific,
Singapore, 1997), which can also be found under cond-mat/9705295.

\bibitem{kardar} For a review of non-equilibrium fluctuations of
manifolds in random media and in particular polymers, flux lines,
and interfaces, see M. Kardar, ``Nonequilibrium dynamics of
interfaces and lines,'' Phys. Rep. {\bf 301}, 85--112 (1998) and
references therein.

\bibitem{fisher} For a review of transport of manifolds in random
media, and in particular charge density waves, flux lines, cracks
and faults, see D. S. Fisher, ``Collective transport in random
media: from superconductors to earthquakes,'' Phys. Rep. {\bf 301},
113--150 (1998) and references therein.

\bibitem{schneider} T. Schneider and E. Pytte, ``Random-field
instability of the ferromagnetic state,'' Phys.
Rev. B {\bf 15}, 1519--1522 (1977).

\bibitem{aharony} A. Aharony, ``Tricritical points in systems
with random fields,'' Phys. Rev. B {\bf 18}, 3318--3327 (1978).

\bibitem{dahmen1} K. Dahmen and J. P. Sethna, ``Hysteresis
loop critical exponents in $6-\epsilon$ dimensions,'' Phys. Rev.
Lett. {\bf 71}, 3222--3225 (1993).

\bibitem{dahmen2} K. Dahmen and J. P. Sethna, ``Hysteresis,
avalanches, and disorder-induced critical scaling: a
renormalization-group approach,'' Phys. Rev. B {\bf 53},
14872--14905 (1996).

\bibitem{silveira} R. da Silveira and M. Kardar, ``Critical
hysteresis for $n$-component magnets,'' Phys. Rev. E {\bf 59},
1355--1367 (1999).

\bibitem{landau} An effective theory for magnetic systems was first
introduced in L. D. Landau, Phys. Z. Sowjetunion {\bf 11}, 26
(1937), reprinted in Collected Papers of L. D. Landau, ed. D. ter
Haar (Pergamon, London, 1965), p. 193, and V. L. Ginzburg and L. D.
Landau, Zh. Eksp. Theor. Fiz. {\bf 20}, 1064 (1950). For a review,
see Ref.~\ref{bib:ma}, Chapter 2 and Ref.~\ref{bib:huang}, Chapter
17.

\bibitem{ma} S.-K. Ma, {\sl Modern Theory of Critical Phenomena} 
(Addison-Wesley Publishing Company, Reading, MA,
1976).~\label{bib:ma}

\bibitem{rg} For an introduction to the renormalization-group
theory, see Ref.~\ref{bib:huang}, Chapter 18,
Ref.~\ref{bib:parisi}, Chapters 7--9. Other good books on the
subject include Ref.~\ref{bib:ma}
and D. J. Amit, {\sl Field Theory, the Renormalization Group,
and Critical Phenomena} (World Scientific, Singapore, 1984), 2nd
ed.

\bibitem{barabasi} A.-L. Barab\'asi and H. E. Stanley, {\sl Fractal
Concepts in Surface Growth} (Cambridge University Press, Cambridge,
1995).

\bibitem{de_gennes} P.-G. de Gennes, ``Wetting: statics and 
dynamics,'' Rev. Mod. Phys. {\bf 57}, 827--863 (1985).

\bibitem{ertas} D. Ertas and M. Kardar, ``Critical dynamics of
contact line depinning,'' Phys. Rev. E {\bf 49}, R2532--2535 (1994).

\bibitem{mandelbrot} B. B. Mandelbrot, {\sl Fractals and Scaling
in Finance: Discontinuity, Concentration, Risk} (Springer Verlag,
1997).

\bibitem{bouchaud} J.-P. Bouchaud and M. Potters,
{\sl Th\'eorie des Risques Financiers} (Al\'ea Saclay, Paris, 1997);
{\sl Theory of Financial Risk}, to be published. 

\bibitem{huberman} B. A. Huberman and R. M. Lukose,
``Social dilemmas and internet congestion,'' Science {\bf 277},
535--537 (1997) and references therein.

\bibitem{dynamics} For a review of dynamical critical phenomena, see
P. C. Hohenberg and B. I. Halperin, ``Theory of dynamical critical
phenomena,'' Rev. Mod. Phys {\bf 49}, 435--479 (1977) and
Ref.~\ref{bib:ma}, Chapters 11--14.

\end{thebibliography}
\end{document}